\documentclass{pasj02}
\usepackage[switch,mathlines]{lineno}
\usepackage{url}
\usepackage{natbib}
\usepackage{comment}

\jyear{2026}
\Received{2026/6/19}%{yyyy/mm/dd}
\Accepted{2026/9/8}%{yyyy/mm/dd}
\begin{document}
\title{Non-Thermal Pressure due to Gas Motions in the Intracluster Medium: Confronting XRISM/Resolve with TNG-Cluster Simulations}
%%% begin:list of authors
% Do NOT capitalize all letters in "textsc".
\author{
 Erwin T. \textsc{Lau},\altaffilmark{1}\altemailmark\orcid{0000-0001-8914-8885} \email{erwin.lau@cc.nara-wu.ac.jp}
 Naomi \textsc{Ota},\altaffilmark{1,2}\orcid{0000-0002-2784-3652}
 and
 Daisuke \textsc{Nagai}\altaffilmark{3}\orcid{0000-0002-6766-5942}
}
\altaffiltext{1}{Department of Physics, Nara Women’s University, Kitauoyanishi-machi, Nara, Nara 630-8506, Japan}
\altaffiltext{2}{Argelander-Institut f\"{u}r Astronomie (AIfA), Universit\"{a}t Bonn, Auf dem H\"{u}gel 71, 53121 Bonn, Germany}
\altaffiltext{3}{Department of Physics, Yale University, New Haven, CT 06520, USA}

\KeyWords{galaxies: clusters --- intergalactic medium --- X-rays: galaxies: clusters --- methods: numerical}

\maketitle

\begin{abstract}
Intracluster medium (ICM) gas motions probe cluster assembly, feedback, and non-thermal pressure support, but recent XRISM observations reveal velocity dispersions and non-thermal pressure fractions systematically lower than simulations predict, with extreme systems such as Abell~2029 falling below nearly all simulated clusters. Using the TNG-Cluster simulations, we show that the non-thermal pressure fraction depends sensitively on cool-core state and formation history, and provide a two-scale fitting function capturing both the inner cool-core suppression and outer rise of the radial profile. By forward-modeling mock XRISM observations and comparing them with both projected and intrinsic three-dimensional quantities, we find that azimuthal variations and projection effects contribute to the deficit in the observed velocity dispersion and non-thermal pressure fraction. This bias increases with radius and partially offsets the intrinsic outward rise in the true three-dimensional non-thermal pressure fraction. However, these effects cannot explain the extremely low values of the non-thermal pressure fraction observed in Abell 2029, which fall below approximately the 0th-6th percentiles of the simulated cool-core cluster distribution at every measured radius under both the turbulence-only and turbulence-plus-bulk definitions.  The remaining tension points to rare dynamical conditions or missing physics affecting the amplitude of gas motions in current ICM models.
\end{abstract}

%\pagewiselinenumbers

\section{Introduction}

Galaxy clusters, the most massive gravitationally bound systems in the universe, form through hierarchical mergers and continuous accretion from the cosmic web. A fraction of the gravitational energy released during structure formation is converted into kinetic energy of the intracluster medium (ICM), generating bulk flows and turbulence that can persist over cosmological timescales \citep{rasia_etal04, dolag_etal05, nagai_etal07b}. These gas motions contribute to non-thermal pressure support \citep{lau_etal09, vazza_etal09} that impacts hydrostatic mass estimates \citep{rasia_etal06, nagai_etal07b, lau_etal13, Biffi:2016, Gianfagna2023} and provides insight into cluster assembly \citep{nelson_etal12, nelson_etal14} and feedback processes \citep{Yang:2016, Bourne2017, lau_etal17, Li2020}. We refer the reader to \citet{simionescu_etal19} for a review of the ICM gas motions and their observational signatures.

Cosmological hydrodynamical simulations predict that non-thermal pressure is a non-negligible component of the ICM, increasing with radius as clusters transition from relatively relaxed cores to dynamically active outskirts. Typical models find non-thermal pressure fractions of $\lesssim 10\%$ in central regions, rising to $\gtrsim 20$--$30\%$ near the virial radius \citep{nagai_etal13, nelson_etal14b, vazza_etal17, angelinelli_etal20}.

High-resolution X-ray spectroscopy has recently enabled direct measurements of ICM gas motions. The {Hitomi} observation of the Perseus cluster revealed unexpectedly low velocity dispersions in the core \citep{hitomi_nature_2016, hitomi_two_t_2018}, suggesting a relatively quiescent ICM. More recently, XRISM/Resolve has extended such measurements to a larger sample of clusters and to larger radii, providing the first systematic constraints on ICM kinematics across different dynamical regimes \citep[e.g.][]{xrism_coma, xrism_a2319, xrism_ophiuchus, xrism_a754_2,xrism_a1795}.

Recent XRISM/Resolve results indicate that observed gas motions are systematically lower than predicted by current simulations. In particular, velocity dispersions in cool-core regions are below the median predictions of multiple simulation suites by factors of $\sim 1.5$--$1.7$ \citep{xrism_comparison}, corresponding to significantly suppressed non-thermal pressure fractions. Extreme cases are also observed at intermediate radii, beyond the cool core: the relaxed cluster Abell~2029, observed along the northern direction out to $r\simeq0.4\,R_{500c}$ ($\sim R_{2500c}$), exhibits exceptionally low velocity dispersions and non-thermal pressure support ($<2\%$), values not reproduced in existing simulations \citep{xrism_a2029_2}. These discrepancies suggest either limitations in current models of ICM physics or the influence of observational effects such as projection and spatial sampling. Simulation studies aimed at interpreting XRISM/Resolve observations suggest that multiple dynamical processes, including AGN feedback, gas sloshing, and buoyancy-driven motions, contribute to the observed velocity structure of cool-core clusters, and that no single mechanism appears sufficient to explain XRISM/Resolve observations of Perseus \citep{Bellomi2025}.

Interpreting these results requires a direct connection between simulations and observations. In particular, it is necessary to quantify how intrinsic three-dimensional gas motions map onto line-of-sight measurements obtained with finite spatial coverage. Azimuthal variations and projection effects may introduce substantial scatter in inferred quantities, especially for spatially limited observations \citep{ota_etal18}. Recent numerical modeling of the XRISM/Resolve velocity structure in Coma further shows that moderate velocity dispersions can arise along with coherent merger-driven bulk flows, motivating the joint interpretation of line-of-sight velocities and dispersions when analyzing ICM kinematics \citep{Zhang2026}.

In this work, we analyze the TNG-Cluster simulation \citep{Nelson2024}, a suite of high-resolution zoom-in cosmological magnetohydrodynamic simulations, to investigate the origin and observational inference of non-thermal pressure in galaxy clusters. The zoom-in approach enables improved resolution of ICM dynamics in massive clusters, allowing for a detailed characterization of the velocity structure and its dependence on cluster properties. Recent work based on the TNG-Cluster simulation has begun to characterize the velocity structure of the ICM in detail. \citet{Ayromlou2024} presented a comprehensive analysis of gas motions across the cluster population, establishing baseline trends in turbulence, anisotropy, and radial dependence. \citet{Truong2024} performed end-to-end XRISM/Resolve mock observations of Perseus-like clusters, showing that the simulated ICM exhibits subsonic turbulence and low inferred velocity dispersions consistent with {Hitomi} measurements. Most recently, \citet{saha_etal26} applied a multi-scale filtering Reynolds decomposition to separate the intrinsic velocity field into bulk and turbulent components. These studies highlight the importance of projection effects and feedback-driven dynamics in shaping observable gas motions, but they do not fully explore the implications for the broader XRISM/Resolve cluster sample or the origin of the observed discrepancies. While previous work has demonstrated that the TNG-Cluster can reproduce low velocity dispersions in individual cool-core systems, comparisons across cluster samples reveal a systematic offset between simulations and XRISM measurements \citep{xrism_comparison}.

Our analysis builds on and complements these works. Whereas previous studies characterize the intrinsic three-dimensional velocity field, either in full \citep{Ayromlou2024} or decomposed into bulk and turbulent components \citep{saha_etal26}, we forward-model the full XRISM/Resolve observable, following the intrinsic motions through emission weighting, instrumental response, projection, and azimuthal sampling, and confront the result directly with the measurements of North arm of Abell~2029 (A2029N). We further characterize how the non-thermal pressure fraction depends on the integrated formation history of the cluster, and provide a calibrated fitting function for its radial profile across cool-core states. In doing so, we connect the intrinsic velocity structure described by these earlier studies to the observational biases that govern its inference, quantifying the extent to which projection and azimuthal sampling can or cannot account for the low gas motions observed by XRISM.

The paper is organized as follows. In Section~\ref{sec:method} we describe the simulation data and analysis methods. In Section~\ref{sec:results} we present the results, including three-dimensional profiles, fitting models, and mock XRISM analyses. In Section~\ref{sec:discussions} we discuss limitations of current work and future directions, and in Section~\ref{sec:summary} we summarize our conclusions.

\section{Methodology}
\label{sec:method}

\subsection{The TNG-Cluster Simulation}\label{sec:tng_cluster_sim}

We make use of the publicly available TNG-Cluster simulation \citep{Nelson2024} to study non-thermal pressure support in massive galaxy clusters. TNG-Cluster is a suite of zoom-in cosmological magnetohydrodynamical simulations based on the IllustrisTNG model, designed to follow the formation and evolution of individual massive galaxy clusters with high resolution. The simulations are evolved with the moving-mesh code \textsc{AREPO} \citep{arepo}, and include a comprehensive galaxy formation model with radiative cooling, star formation, chemical enrichment, and feedback from supernovae and active galactic nuclei (AGN).

Compared to large-volume simulations, the TNG-Cluster zoom-in approach enables improved resolution of the intracluster medium (ICM), allowing for a more detailed characterization of gas dynamics, including turbulence, bulk flows, and rotational motions. The simulated clusters span a representative range of masses and dynamical states, making the sample well suited for studying the diversity of ICM velocity structure and non-thermal pressure support.

The simulations provide snapshots across cosmic time, and in this work we focus on the $z=0$ snapshots. Dark matter halos are identified using a friends-of-friends (FoF) algorithm, and gravitationally bound substructures are further identified with the \textsc{SUBFIND} algorithm. Gas properties are extracted directly from the simulation cells, enabling the decomposition of the total pressure into thermal and non-thermal components, with the latter arising from resolved gas motions.

For each of the 352 clusters in the simulation, we construct radial profiles of thermodynamic and kinematic quantities within spherical overdensity radii (e.g., $R_{500c}$ and $R_{200m}$, defined as the radii within which the mean enclosed density is 500 times the critical density and 200 times the mean matter density of the Universe, respectively). This approach allows for a systematic analysis of the non-thermal pressure fraction as a function of radius, and facilitates direct comparison with observational measurements and previous simulation studies.

\subsection{Cool-core classification}\label{sec:cc_classification}

A key aspect of our analysis is the separation of clusters into cool-core (CC), weak cool-core (WCC), and non-cool-core (NCC) systems, as these regimes are governed by different physical processes. In CC systems, the central ICM is strongly influenced by radiative cooling and feedback from the central supermassive black hole (SMBH), whereas in NCC systems and at large radii, gas dynamics are primarily driven by mergers and large-scale structure formation.

To enable a consistent comparison with XRISM observations and recent simulation studies, the simulated clusters are classified according to their central cooling time, defined as
\begin{equation}
t_{\rm cool} = \frac{3}{2} \frac{(n_e + n_i) k_B T}{n_e n_i \Lambda},
\end{equation}
where $n_e$ and $n_i$ are the electron and ion number densities, $T$ is the gas temperature, $\Lambda$ is the cooling function, and $k_B$ is the Boltzmann constant. The cooling time is computed using emission-weighted gas properties within a central aperture of $0.015\,R_{500}$ \citep{Lehle2024}, following observationally motivated definitions \citep{McDonald2013}.

\begin{figure}
    \begin{center}
    \includegraphics[width=0.45\textwidth]{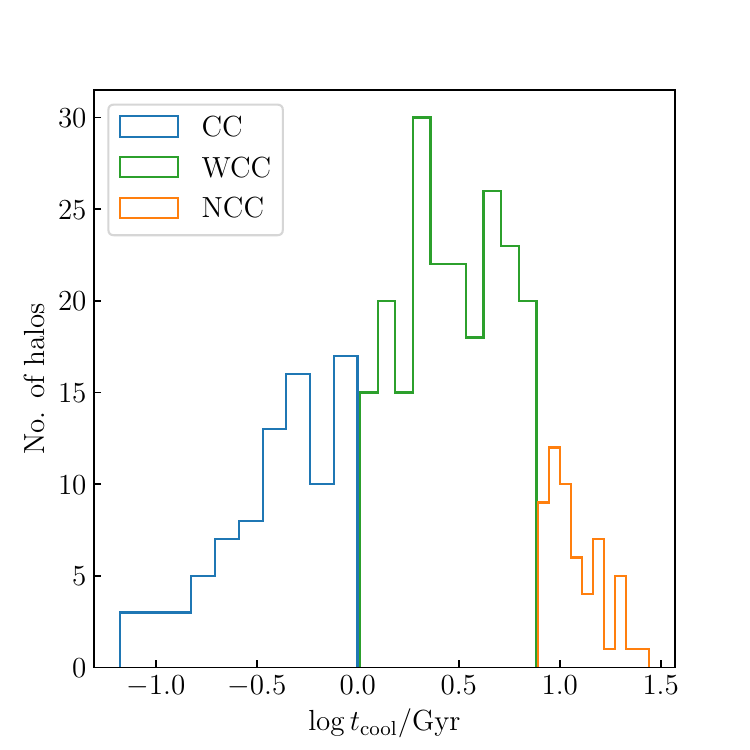}
    \end{center}
    \caption{Distribution of cool-core (CC), weak cool-core (WCC), and non-cool-core (NCC) clusters in the TNG-Cluster simulation sample, as a function of $\log$ of the central cooling time $t_{\rm cool}$ measured within $0.015R_{500c}$ of each halo. 
    %{Alt text: Histogram of the distribution of central cooling times for 352 TNG-Cluster halos. Clusters are classified into cool-core (CC, blue), weak cool-core (WCC, green), and non-cool-core (NCC, orange) systems according to their cooling time measured within 0.015 R500c.}
    }
    \label{fig:cc_ncc_hist}
\end{figure}

Following \citet{Lehle2024}, we adopt the threshold of $t_{\rm cool}/\mathrm{Gyr} < 1$ to identify CC systems, consistent with common observational classifications and recent XRISM simulation comparisons \citep{xrism_comparison}. This threshold value approximately corresponds to clusters in which radiative losses are sufficiently strong to require sustained feedback heating, and therefore where SMBH-driven processes are expected to significantly impact the ICM thermodynamic and kinematic structure. Systems with $t_{\rm cool}/\mathrm{Gyr} \in [1.0, 7.7)$ are classified as WCC, and systems with $t_{\rm cool}/\mathrm{Gyr} \geq 7.7$ are classified as NCC. This results in 85 CC clusters, 211 WCC clusters, and 56 NCC clusters. Figure~\ref{fig:cc_ncc_hist} shows the distributions of the CC/WCC/NCC clusters.

This classification allows us to interpret XRISM measurements in two physically distinct regimes: (i) a feedback-dominated regime, corresponding to CC cores and regions within the cooling radius, and (ii) a gravity-dominated regime, corresponding to NCC systems and regions outside cool cores. This separation is essential for disentangling the effects of AGN feedback from those of hierarchical accretion when comparing simulations to observations. We also note that the cooling time has previously been identified as a principal axis of the X-ray fundamental plane and shown to correlate with ICM morphology through multivariate analyses of X-ray observables \citep{Ota2006}, further supporting the cooling time as a physically relevant classification parameter.

To assess the robustness of our results to the choice of classification metric, we also consider an alternative definition based on central entropy, $K = k_B T n_e^{-2/3}$, computed within the same aperture. We find that adopting an entropy-based threshold yields consistent CC/NCC classifications for the vast majority of systems, and does not affect any of our main conclusions. This agreement reflects the tight correlation between cooling time and entropy in cluster cores.

Finally, we note that the classification is applied consistently between simulations and observations in a statistical sense, but individual XRISM pointings may probe regions that transition between these regimes. In particular, offset pointings in cool-core clusters \citep{xrism_a2029_2} can sample gas outside the cooling radius, effectively probing gravity-dominated conditions despite originating from a CC system. We therefore interpret our results primarily in terms of local physical regime (cool-core vs. outskirts), rather than global cluster classification alone.

\subsection{Profile Calculations}
Radial profiles are constructed by selecting gas cells within a three-dimensional spherical region in 40 logarithmically spaced bins spanning $r \in [1, 10^4]\,\mathrm{kpc}$ from the halo center, with a temperature cut $T \ge 10^6\,\mathrm{K}$ to isolate the hot X-ray emitting intracluster medium. The gas cells are binned in radius using logarithmically spaced spherical shells, such that each bin covers a fixed interval in $\log r$. Within each radial bin, all quantities are computed using mass-weighted averages over the contributing gas cells.

We follow \citet{lau_etal09} to calculate the mass-weighted velocity profiles. The velocity field is decomposed into radial and tangential components relative to the halo center. The mean radial velocity in a given shell is defined as
\begin{equation}
\langle v_r \rangle = \frac{\sum_i w_i\, v_{r,i}}{\sum_i w_i},
\end{equation}
where $w_i$ and $v_{r,i}$ are the weights (mass or emission weighting; see Section~\ref{sec:weightings}) and radial velocity of the $i$-th gas cell. The radial velocity dispersion is then computed as
\begin{equation}
\sigma_r^2 = \frac{\sum_i w_i\, (v_{r,i} - \langle v_r \rangle)^2}{\sum_i w_i}.
\end{equation}
For the tangential motion, a rotation axis is defined locally within each radial bin from the angular momentum of the gas contained in that spherical shell, and velocities are projected onto the corresponding azimuthal direction; because this axis is recomputed shell by shell rather than fixed globally (e.g., within $R_{500c}$ or $R_{200c}$), it can vary in orientation with radius, tracking any radius-dependent change in the angular momentum axis of the gas distribution rather than assuming a single, cluster-wide rotation axis. The mean rotational velocity is given by
\begin{equation}
v_{\rm rot} = \langle v_{\phi} \rangle = \frac{\sum_i w_i\, v_{{\phi},i}}{\sum_i w_i},
\end{equation}
and the mean bulk velocity in 1D is then given by
\begin{equation}
v_{\rm bulk, 1D} = \sqrt{\frac{\langle v_r \rangle^2 + \langle v_{\phi} \rangle^2 + \langle v_{\theta} \rangle^2}{3}}.
\end{equation}
The dispersions in the $\phi$ and $\theta$ directions are
\begin{eqnarray}
\sigma_\phi^2 &=& \frac{\sum_i w_i\, (v_{\phi,i} - \langle v_\phi \rangle)^2}{\sum_i w_i}, \\
\sigma_\theta^2 &=& \frac{\sum_i w_i\, (v_{\theta,i} - \langle v_\theta \rangle)^2}{\sum_i w_i}.
\end{eqnarray}
The total one-dimensional velocity dispersion is obtained by combining the radial and tangential dispersions. The tangential one-dimensional dispersion is then
\begin{equation}
\sigma_{t,\mathrm{1D}}^2=\frac{\sigma_{\phi}^2+\sigma_{\theta}^2}{2},
\end{equation}
so the velocity anisotropy profile is
\begin{equation}
\beta(r)=1-\frac{\sigma_{t,\mathrm{1D}}^2(r)}{\sigma_r^2(r)}.
\end{equation}

We note that our decomposition of the velocity field into a shell-mean (bulk) component and a residual dispersion (turbulent) component differs from the multi-scale filtering Reynolds decomposition \citep{vazza_etal12, vazza_etal17, saha_etal26}, in which the bulk--turbulent separation scale is determined locally and adaptively for each gas cell. For the purpose of comparing to XRISM/Resolve measurements that integrate over a $3'\times3'$ field of view, the shell-based dispersion is the better-matched quantity: a spatially integrated X-ray line width encodes the full velocity variance within the aperture regardless of coherence scale. Both approaches share an irreducible ambiguity in the definition of turbulence, set by the convergence tolerance and filtering length in the multi-scale case and by the bin width in ours; we therefore report both mass-weighted and emission-weighted profiles (Appendix) to bracket the weighting dependence.

Following \cite{lau_etal09}, the non-thermal pressure profile is inferred from the velocity dispersion as
\begin{equation}
P_{\mathrm{turb}}(r) = \frac{1}{3}\rho(r)\sigma_{\mathrm{3D}}^2(r) = \rho(r)\sigma_{\mathrm{1D}}^2(r),
\end{equation}
where $\rho(r)$ is the gas density and $\sigma_{\mathrm{1D}}$ is the mass-weighted one-dimensional velocity dispersion,
\begin{equation}
\sigma_{\mathrm{1D}}^2 = \frac{1}{3}\sigma_{\mathrm{3D}}^2 = \frac{1}{3}\left(\sigma_r^2 +\sigma_\phi^2+\sigma_\theta^2 \right).
\label{eq:sigma_1d}
\end{equation}
Note that this definition of non-thermal pressure differs by a factor of $3$ from that in the A2029N XRISM analysis, where the non-thermal pressure is defined as $P_{\mathrm{turb}}(r) =\rho\sigma_{\mathrm{3D}}^2 = 3\rho\sigma_{\mathrm{1D}}^2$. In our definition, the non-thermal pressure includes only the contribution from turbulent gas motions and explicitly excludes coherent bulk flows. The total pressure support is then defined as
\begin{equation}
P_{\rm tot} = P_{\mathrm{turb}} + P_{\mathrm{thermal}},
\end{equation}
where $P_{\mathrm{thermal}}$ is the thermal pressure, and the non-thermal pressure fraction is
\begin{equation}
f_{\rm nth} \equiv P_{\mathrm{turb}}/P_{\rm tot}.
\end{equation}
Another way of defining the non-thermal pressure fraction is to include the bulk motions $v_{\rm bulk}$ in the non-thermal pressure:
\begin{equation}
P_{\mathrm{turb+bulk}}(r) = \rho(r)\left(\sigma_{\mathrm{1D}}^2 + v_{\rm bulk, 1D}^2 \right),
\end{equation}
with the non-thermal pressure fraction as
\begin{equation}
f_{\rm nth}^{\rm turb+bulk} \equiv P_{\mathrm{turb+bulk}}/P_{\rm tot}.
\end{equation}
Unless stated otherwise, whenever we mention non-thermal pressure and non-thermal pressure fraction, we refer to the definitions excluding bulk motions.

\subsection{X-ray Emission Weightings}\label{sec:weightings}

To enable comparison between simulation predictions and X-ray spectroscopic measurements, particularly those anticipated from XRISM/Resolve, we computed emission-weighted kinematic profiles alongside the standard mass-weighted profiles. For mass-weighted profiles, the weight is the gas mass of the Voronoi cell. For emission weighting, following the convention of \citet{rasia_etal05}, the weight is computed as
\begin{equation}
    w_i = n_{e,i}^{2}\,V_i\,\Lambda(T_i,\, Z_i),
    \label{eq:em_weight}
\end{equation}
where $n_{e,i}$ is the electron number density, $V_i$ is the Voronoi cell volume, and $\Lambda(T_i, Z_i)$ is the X-ray emissivity per unit emission measure evaluated at cell temperature $T_i$ and metallicity $Z_i$. We refer to this scheme as ``emission weighting'' (rather than ``spectroscopic weighting,'' since Equation~\ref{eq:em_weight} involves only the integrated emissivity and does not incorporate the spectral-shape information, e.g., line widths from a multi-temperature, multi-velocity gas column, that a true spectroscopic weighting would require). The emissivity function $\Lambda(T, Z)$, generated with \textsc{pyAtomDB}\footnote{\url{https://github.com/AtomDB/pyatomdb}} under the assumption of collisional ionization equilibrium, is interpolated from a precomputed table, integrated over the $0.1$--$10\,\mathrm{keV}$ band and convolved with the XRISM/Resolve spectral response matrix (RMF) \verb|rsl_Hp_5eV.rmf| and ancillary response file (ARF) \verb|rsl_pointsource_GVclosed.arf| used in generating the XRISM/Resolve mocks, described in Section~\ref{sec:mock_data}.

\begin{figure*}
    \begin{center}
    \includegraphics[width=0.99\textwidth]{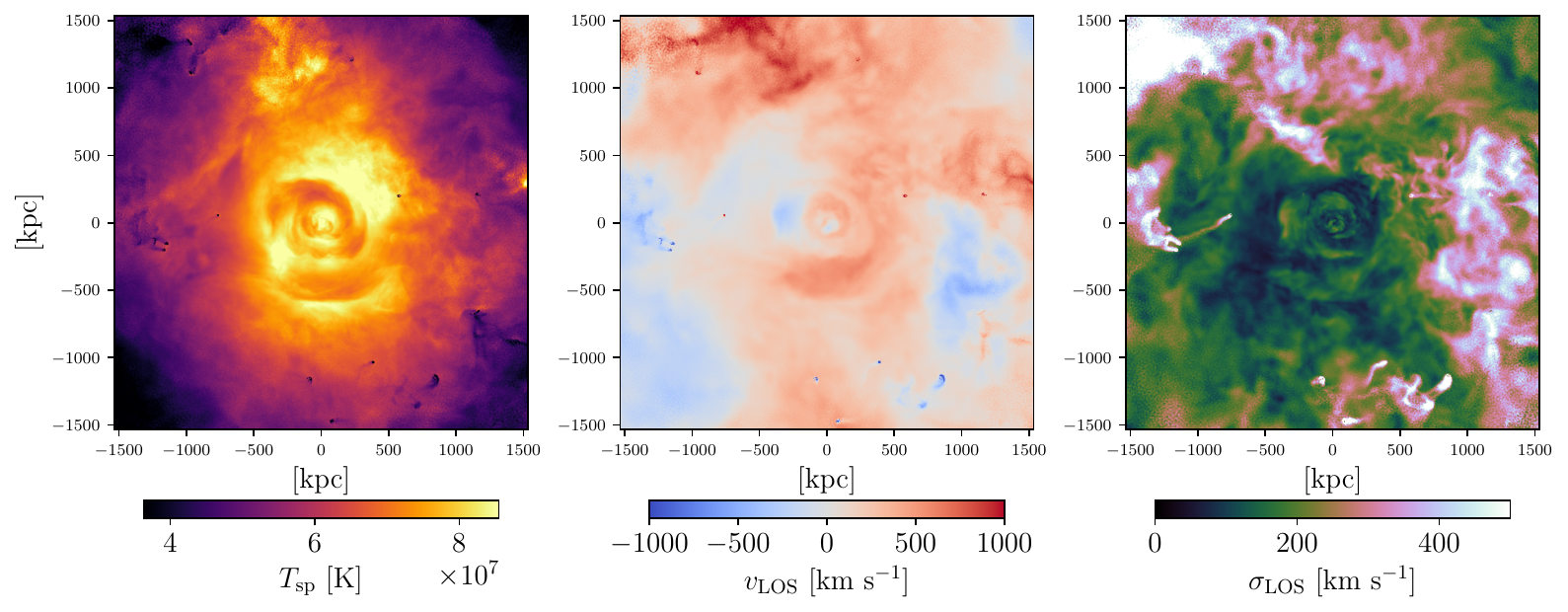}
    \end{center}
    \caption{X-ray emission-weighted projected maps of a representative cool-core cluster (HaloID 20) projected along the x-axis. From left to right, the panels show the X-ray emission-weighted temperature $T_{\rm sp}$, line-of-sight bulk velocity $v_{\rm LOS}$, and line-of-sight velocity dispersion $\sigma_{\rm LOS}$. The maps span $3\times 3$ Mpc centered on the cluster, with color bars indicating the ranges of temperature, LOS velocity, and LOS velocity dispersion.
    %{Alt text: Three projected maps of a simulated cool-core galaxy cluster viewed along the x-axis. The left panel shows X-ray emission-weighted temperature, the middle panel shows line-of-sight velocity, the right panel shows line-of-sight velocity dispersion.}
    }
    \label{fig:projected_maps}
\end{figure*}

\begin{figure}
    \begin{center}
    \includegraphics[width=0.48\textwidth]{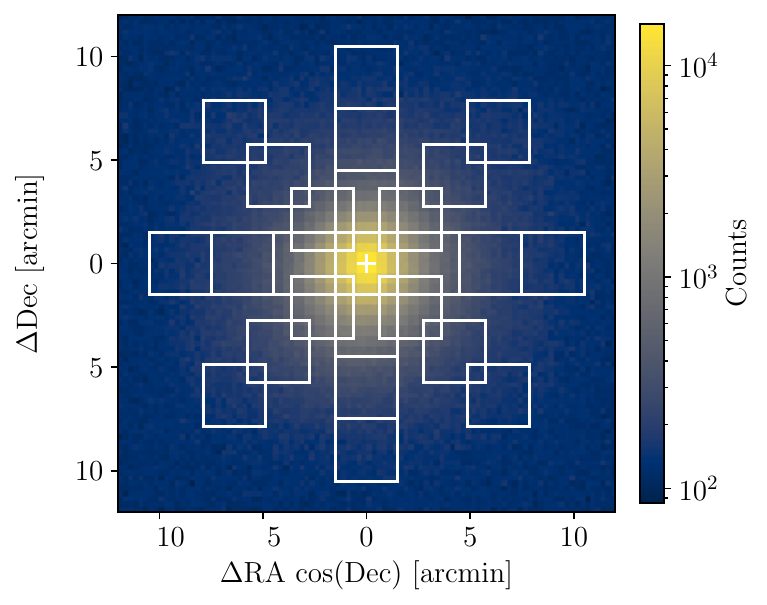}
    \end{center}
    \caption{Mock XRISM/Resolve map with 1Ms exposure of the projection along the x-axis for a halo in TNG-Cluster with HaloID 20, a CC cluster in the TNG-Cluster simulation (the same projected cluster as in Figure~\ref{fig:projected_maps}). Each white square indicates the $3'\times 3'$ (approximately $270$ kpc $\times$ $270$ kpc at the observed redshift $z=0.0787$) FOV of each pointing. The color indicates the photon counts.  
    %{Alt text: Mock XRISM X-ray image of a simulated cool-core cluster projected along the x-axis. White squares indicate the XRISM/Resolve fields of view arranged in eight azimuthal directions and four radial pointings per direction, illustrating the observational sampling strategy used for the mock observations.}
    }
    \label{fig:xrism_mock_map}
\end{figure}

\subsection{Projected Maps}\label{sec:proj_maps}

We generate mass-weighted and X-ray emission weighted projected maps in temperature, metallicity, and line-of-sight bulk velocity and velocity dispersion. To construct the projected maps, we first deposit the hot gas cells ($T>10^{6}\,\mathrm{K}$) onto a uniform Cartesian grid centered on each halo (a $512^{3}$ mesh with $6\,\mathrm{kpc}$ cells), accumulating the gas density, the mass-weighted temperature, the mass-weighted metallicity, and the three mass-weighted velocity components in every voxel via a cloud-in-cell assignment. For each of the three orthogonal lines of sight (the $x$, $y$, and $z$ axes of the simulation box) we then collapse the cube along the projection direction using either the total gas mass as weight, or the emission weight in Equation~(\ref{eq:em_weight}) which approximates the X-ray emissivity of each voxel. The weighted line-of-sight bulk velocity, velocity dispersion, and temperature in a given map pixel are
\begin{eqnarray}
V_{\rm bulk}&=&\frac{\sum_i w_i v_{i}}{\sum_i w_i}, \\
\sigma_{}^{2}&=&\frac{\sum_i w_i\,(v_{i}-V_{\rm bulk})^{2}}{\sum_i w_i}, \\
T&=&\frac{\sum_i w_i T_i}{\sum_i w_i},
\label{eq:projmaps}
\end{eqnarray}
where the sums are applied to all voxels along the line of sight within that pixel, $v_i$ is the velocity component in voxel $i$ along the projection axis, and $w_i$ is the weight in voxel $i$. Repeating this for every pixel yields two-dimensional maps of $V_{\rm bulk}$, $\sigma$, and $T$, which we sample in eight azimuthal sectors (in the same way as the mock XRISM maps in Section~\ref{sec:mock_data}). Figure~\ref{fig:projected_maps} shows the projected maps for a representative cool-core cluster. The temperature map reveals a disturbed cool-core morphology with a prominent spiral cold-front/sloshing pattern extending several hundred kpc from the cluster center. The LOS velocity map shows coherent large-scale motions associated with the sloshing gas, whereas the velocity-dispersion map highlights enhanced turbulent motions along cold fronts and in the cluster outskirts.

\begin{figure*}
    \begin{center}
    \includegraphics[width=0.99\textwidth]{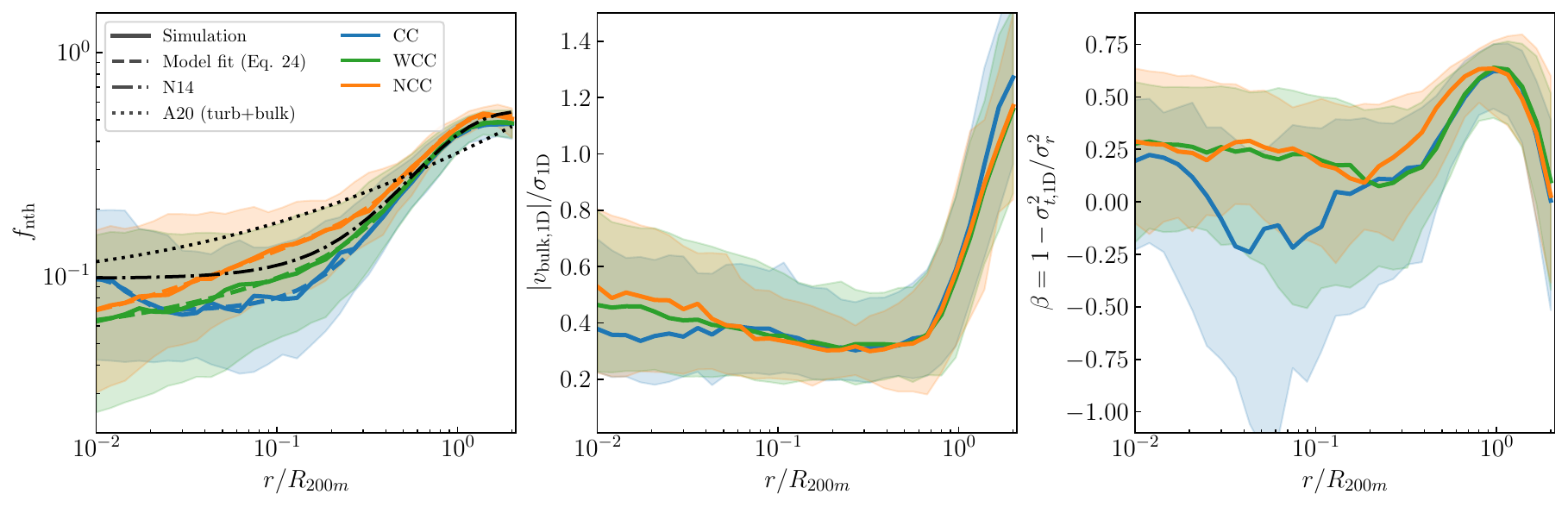}
    \end{center}
    \caption{
        {\em Left} panel: Median profiles of the non-thermal pressure fraction as a function of scaled radius $r/R_{200m}$ in bins of cool-core state: strong cool-core (CC), weak cool-core (WCC), and non-cool-core (NCC), with central cooling times $t_{\rm cool}/{\rm Gyr} \leq 1$, $t_{\rm cool}/{\rm Gyr} \in [1.0, 7.7)$, and $t_{\rm cool}/{\rm Gyr} \geq 7.7$, respectively. The shaded regions indicate 16th--84th percentiles. We also show the best-fit model (Equation~\ref{eq:fnt_model}) for each of the cool-core state bins as dashed lines. The model profiles from \cite{nelson_etal14} (N14, dot-dashed black line) and \cite{angelinelli_etal20} (A20, dotted black line), calibrated using two different cosmological simulations without radiative cooling and feedback, are included for comparison. For the \cite{angelinelli_etal20} model, the non-thermal pressure includes contributions from both bulk and turbulent motions.
        {\em Middle} panel: Median profiles of the ratio of bulk velocity to one-dimensional velocity dispersion, for the same cool-core state bins as in the left panel.
        {\em Right} panel: Velocity anisotropy profiles for the same clusters. Negative values indicate the velocity dispersions are more tangential, while positive values indicate they are more radial.
        %{Alt text: Three-panel comparison of median three-dimensional gas-motion properties for cool-core (CC), weak cool-core (WCC), and non-cool-core (NCC) clusters as a function of radius normalized by R200m. Left: non-thermal pressure fraction. Middle: ratio of bulk velocity to velocity dispersion. Right: velocity anisotropy.}
    }
    \label{fig:fnt_profile_cc}
\end{figure*}

\subsection{Mock XRISM/Resolve Simulations}\label{sec:mock_data}

To compare with the XRISM observations, we generate synthetic XRISM X-ray observations of 3 orthogonal projections for all 352 simulated clusters with the \textsc{pyXSIM} \citep{pyxsim} and \textsc{SOXS} \citep{soxs} packages, which enable forward-modeling of X-ray emission from hydrodynamical simulations and the application of instrumental responses. For each cluster, we construct photon samples from the gas cells using \textsc{pyXSIM}, assuming a collisionally ionized plasma described by the \texttt{APEC} spectral model (version 3.1.2), with thermal line broadening. Gas temperatures, densities, and metallicities are taken directly from the simulation. The emission is generated over the energy range $E \in [0.1,10]$~keV, ensuring full coverage of the XRISM/Resolve bandpass. We fix the redshift of all clusters to $z=0.0787$, the same as Abell~2029. We also fix the Galactic absorption column density to $N_{H}/(10^{22}\,\mathrm{cm^{-2}}) = 0.018$.

The photon lists are then projected along a chosen line of sight and convolved with the instrumental response using \textsc{SOXS}. We adopt the XRISM/Resolve response files corresponding to the closed gate-valve configuration, including both the RMF \verb|rsl_Hp_5eV.rmf| and ARF \verb|rsl_pointsource_GVclosed.arf|. Note that the ARF we used applies only to point sources, so we ignore effects due to the finite point spread function (PSF). This is justified as we extract only a single spectrum from each pointing that covers the entire field of view (FOV). For each pointing, we assume an exposure time of $1000~\mathrm{ks}$ to obtain as much photon statistics as possible. The spatial resolution is set to match that of XRISM/Resolve. 

For each mock map, we extract spectra in eight azimuthal directions. Each direction contains four radial pointings, with each pointing covering the $3'\times 3'$ field of view of XRISM/Resolve.  
Figure~\ref{fig:xrism_mock_map} shows the mock XRISM map of an example cool-core cluster (HaloID 20), selected to be a dynamically relaxed cluster with a sloshing pattern similar to that of A2029. The pointing configuration is shown in the figure.

\section{Results}
\label{sec:results}

\begin{figure*}
    \begin{center}
    \includegraphics[width=0.99\textwidth]{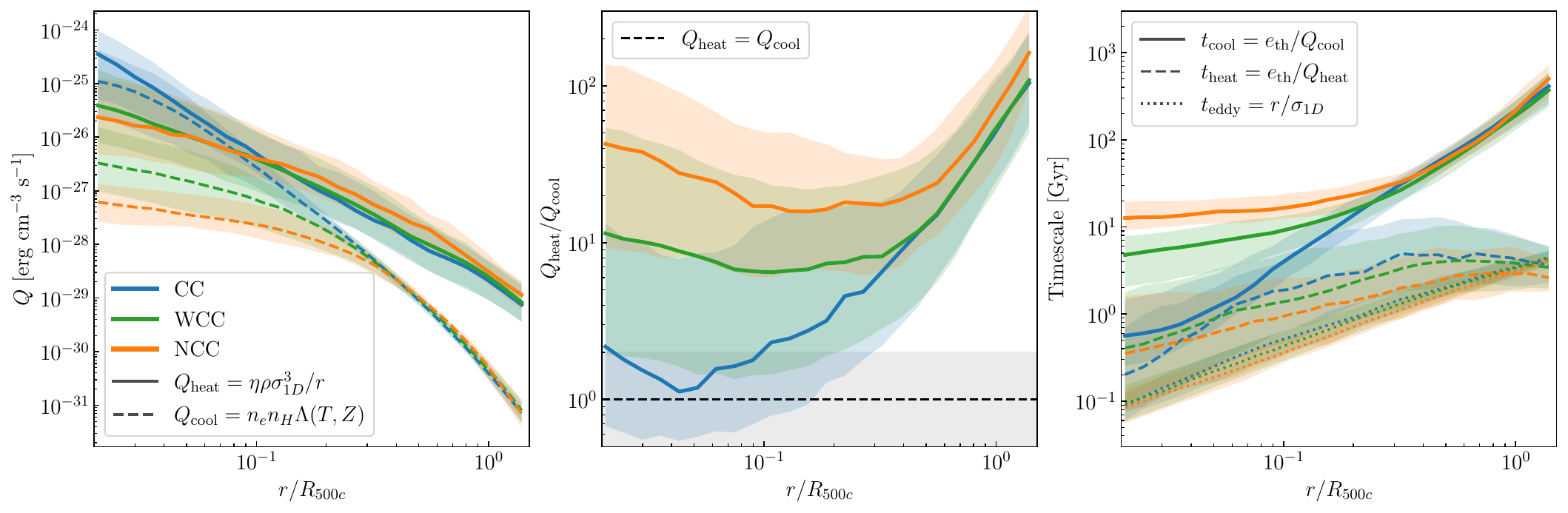}
    \end{center}
    \caption{
        \textit{Left} panel: volumetric turbulent heating rate $Q_{\rm heat}$ (solid) and radiative cooling rate $Q_{\rm cool} = n_e\,n_H\,\Lambda(T,Z)$ (dashed) as a function of scaled radius for the clusters binned by their cool-core states (CC: blue; WCC: green; NCC: orange).
        \textit{Middle} panel: Ratio $Q_{\rm heat}/Q_{\rm cool}$. The horizontal dashed line marks the point where heating exactly balances cooling; the grey band indicates agreement within a factor of two.
        \textit{Right} panel: cooling time $t_{\rm cool} = e_{\rm th}/Q_{\rm cool}$ (solid), turbulent heating time $t_{\rm heat} = e_{\rm th}/Q_{\rm heat}$ (dashed), and eddy turnover time $t_{\rm eddy} = r/\sigma_{1D}$ (dotted) as a function of radius. In all panels, thick curves show the median over all halos in each class; shaded bands denote the 16th--84th percentile scatter.
        {Alt text: Three-panel comparison of turbulent heating and radiative cooling. Left: turbulent heating rate and radiative cooling rate versus radius for CC, WCC, and NCC clusters. Middle: ratio of heating to cooling. Right: cooling, heating, and eddy-turnover timescales.}
    }
    \label{fig:tng_heating_cooling}
\end{figure*}

\subsection{3D Profiles}
\label{sec:3d_profiles}

Unless explicitly stated otherwise, all quantities presented in this Section~\ref{sec:3d_profiles} are mass-weighted; Section~\ref{sec:mock_analyses} explicitly distinguishes mass-weighted (``mw'') and emission-weighted (``em'') quantities wherever both are compared, and labels each accordingly in figure legends and in-text notation.

\subsubsection{Dependence on Cool-core Status}\label{sec:cc_status}

Figure~\ref{fig:fnt_profile_cc} shows the median mass-weighted radial profiles of the non-thermal pressure fraction, velocity anisotropy, and rotational support for clusters separated by cool-core (CC), weak cool-core (WCC), and non-cool-core (NCC) systems.

The radial dependence of the non-thermal pressure fraction, $f_{\rm nth} \equiv P_{\rm turb}/P_{\rm tot}$, differs between the cool-core subsamples. For CC clusters, there is a `V'-shaped trend in $f_{\rm nth}$, where it first decreases from the center to the edge of the core at $\sim 0.3\,R_{200m} \approx 0.1\,R_{500c}$, and then increases at larger radii. At fixed radius, CC clusters exhibit systematically lower non-thermal pressure support in the inner regions compared to NCC systems, while WCC clusters lie in between. This trend reflects the more relaxed dynamical state of CC systems and the enhanced gas motions in NCC clusters associated with recent mergers. The differences among subsamples decrease toward larger radii, where all profiles converge. Beyond the outer accretion shock radius, $\sim 2\,R_{200m}$, $f_{\rm nth}$ for all samples converges to a value of $\sim 0.5$.

Our results are broadly consistent with previous analyses of ICM gas motions \citep{lau_etal09, Biffi:2016, Ayromlou2024}, which also find a systematic increase of non-thermal pressure with radius and significant halo-to-halo scatter. However, by explicitly separating clusters by cool-core state and formation history, we further quantify the physical origin of this scatter.

The velocity anisotropy profile, $\beta(r)$, shows that CC clusters tend to have more tangentially biased motions ($\beta < 0$) in the inner regions, while NCC and WCC systems are mildly radially biased ($\beta > 0$) over the same radial range -- opposite senses of anisotropy. The magnitude of this anisotropy is comparable between the populations at the smallest radii probed, but $|\beta|$ is generally larger for NCC and WCC than for CC over most of $r/R_{200m}\lesssim0.3$--$0.4$: the CC median passes back through $\beta\approx0$ (near isotropy) in this range, whereas NCC and WCC remain steadily anisotropic. Only in a narrow dip near $r\sim0.05$--$0.1\,R_{200m}$ does the CC median reach a magnitude comparable to the steady NCC/WCC value. The scatter in $\beta$ is substantial, and is particularly wide for CC clusters, indicating significant halo-to-halo variation in core kinematics, with individual CC halos reaching much larger $|\beta|$ than the median in this dip.

The ratio of bulk velocity to velocity dispersion, $v_{\rm bulk, 1D}/\sigma_{\rm 1D}$, remains below unity for $r<2\,R_{200m}$, indicating that random motions dominate over bulk motions within the cluster. NCC clusters show slightly enhanced bulk motion in the inner regions compared to CC systems. The $v_{\rm bulk, 1D}/\sigma_{\rm 1D}$ ratio declines with radius for all subsamples until it reaches the outer accretion shock radius, consistent with the dominance of turbulent motions after the infalling gas is shock-heated and the bulk kinetic energy is converted to random gas motions and thermal energy. We note that this result appears to differ from XRISM observations, in which the projected velocity dispersions are much smaller than the projected line-of-sight bulk velocities. These differences arise from projection effects, as we investigate in Section~\ref{sec:mock_analyses}.

\subsubsection{Turbulent heating versus radiative cooling}
\label{sec:heat_cool}

We estimate the turbulent heating rate per unit volume as the rate of kinetic energy dissipation in a Kolmogorov cascade injected at the local radius,
\begin{equation}
    Q_{\rm heat}(r) = \eta\,\rho_{\rm gas}\frac{\sigma_{1D}^3}{r},
    \label{eq:Qheat}
\end{equation}
where $\rho_{\rm gas} = \mu_e\,m_p\,n_e$ is the gas mass density, $\sigma_{1D}$ is the mass-weighted one-dimensional velocity dispersion from the radial profile, $r$ is the three-dimensional radius (taken as a proxy for the turbulent injection scale), and $\eta \approx 5$ is a prefactor calibrated by simulations of hydrodynamic turbulence \citep{zhuravleva_etal14b}, absorbing the geometric factors relating $\sigma_{\rm 1D}$ to the three-dimensional cascade rate. This choice of $\eta$ assumes the turbulence is injected at the local radius $r$, appropriate for large-scale, merger-driven turbulence in which large-scale motions are naturally truncated at $r$ by the finite volume and by gravitational stratification; it is less obviously appropriate for turbulence driven directly on small scales by AGN feedback, where the injection scale is more likely set by the size of the AGN-inflated bubbles or jet lobes than by $r$ itself. Because this scale is not directly measurable in projected X-ray observations, we adopt this estimate throughout as our fiducial estimator, following the same two-injection-scale reasoning discussed by \citet{Zhang_etal26b} for Perseus. Equation~(\ref{eq:Qheat}) follows directly from the turbulent heating rate of \citet{zhuravleva_etal14b}, where $\eta \approx 5$ for a pointwise velocity amplitude at wavenumber $k=2\pi/r$. However, if instead $\sigma_{\rm 1D}$ is interpreted as tracing the cumulative kinetic energy integrated over the full inertial range above $k$, the same Kolmogorov spectrum will give $\eta \approx3$, about half of our assumed value. Given this factor-of-two ambiguity in how $\sigma_{\rm 1D}$ and $r$ map onto the wavenumber-space quantities of \citet{zhuravleva_etal14b}, and the additional uncertainty in the injection scale itself, we regard $\eta$ as uncertain at the factor-of-a-few level, and treat Equation~(\ref{eq:Qheat}) throughout as an order-of-magnitude estimator rather than a precise calculation. 
The radiative cooling rate per unit volume is computed as
\begin{equation}
    Q_{\rm cool}(r) = n_e n_H\Lambda(T,\,Z),
    \label{eq:Qcool}
\end{equation}
where $n_H = 0.83\,n_e$ for a fully ionized plasma with cosmic abundances, and $\Lambda(T,Z)$ is the bolometric cooling function computed with the \textsc{APEC} code. The isochoric thermal energy density $e_{\rm th} = (3/2)\,n_{\rm gas}\,k_BT$ defines the cooling time $t_{\rm cool} = e_{\rm th}/Q_{\rm cool}$, the turbulent heating time $t_{\rm heat} = e_{\rm th}/Q_{\rm heat}$, and the eddy turnover time $t_{\rm eddy} = r/\sigma_{\rm 1D}$.

Figure~\ref{fig:tng_heating_cooling} shows that the turbulent heating budget is strongly cool-core-dependent. For WCC and NCC clusters, $Q_{\rm heat}/Q_{\rm cool}$ is well above unity at every radius shown, of order $5$--$10$ for WCC and $15$--$50$ for NCC even at the smallest radii probed, so turbulent heating alone is comfortably sufficient to offset cooling throughout these systems. For CC clusters, the ratio is close to, but not clearly above, unity: it declines from $\sim2$ near the cluster center to a minimum of $\sim1.1$--$1.2$ at $r\simeq0.05\,R_{500c}$ -- within the grey band marking agreement to within a factor of two -- before rising again at larger radii. This does not imply that merger-driven turbulence, at the large injection scale assumed here, is by itself sufficient in CC cores, even before considering the additional, more centrally concentrated contribution expected from AGN-driven turbulence injected on smaller scales. We caution against over-interpreting even the CC result as demonstrating that turbulent dissipation alone balances cooling: the median sits at the boundary of rough parity, the halo-to-halo scatter (shaded bands) spans values both above and below unity at the same radius, and given the factor-of-a-few uncertainty in $\eta$ discussed above, the true ratio could plausibly sit below unity rather than at parity. The result indicates only that a global cooling catastrophe is not obviously required by this order-of-magnitude energy budget in CC cores, and this conclusion holds more robustly in WCC/NCC systems, where $Q_{\rm heat}/Q_{\rm cool}$ remains well above unity even allowing for the same uncertainty in $\eta$, consistent with the well-established result that some combination of turbulent and AGN-feedback-driven heating offsets cooling in cool cores \citep[e.g.][]{McNamara:2007, Fabian:2012, McNamara_etal26}, rather than requiring AGN feedback as a wholly separate channel from turbulence. It is WCC and NCC clusters that show the clearest, least ambiguous heating dominance. Beyond the dip near $r\sim0.05$--$0.1\,R_{500c}$, the ratio rises steadily outward for all three classes, reaching values of order $10$--$100$ by $r\sim1\,R_{500c}$: at these larger radii $Q_{\rm cool}$ has dropped by several orders of magnitude (owing to the steep $n_e^2$ dependence), while $Q_{\rm heat}$ declines more gently ($\propto \rho\,\sigma^3/r$, with the velocity dispersion remaining substantial beyond $R_{500c}$).

The right panel of Figure~\ref{fig:tng_heating_cooling} translates these rates into timescales: the cooling time in CC cores is $t_{\rm cool} \sim 0.1$--$1\,\rm Gyr$, below the Hubble time $\sim 10\,{\rm Gyr}$, and the turbulent heating time $t_{\rm heat}$ is now comparable to, rather than an order of magnitude longer than, $t_{\rm cool}$ in the same region. The eddy turnover time $t_{\rm eddy} = r/\sigma_{\rm 1D}$ remains smaller than both timescales by 1--2 orders of magnitude, indicating that many eddy turnovers still occur within a cooling time, consistent with turbulent dissipation being a physically plausible, though not uniquely required, heating channel. 

This heating budget does not diminish the direct evidence, from \citet{Prunier2025}, who also analyze the same TNG-Cluster simulations, that AGN-driven weak shocks, detected in 30\% of the TNG-Cluster halos within ${\sim}100\,\rm kpc$ of the central SMBH, carry powers comparable to the cooling luminosity ($10^{44\text{--}46}\,\rm erg\,s^{-1}$): the dominant core heating channel identified directly in the simulations is shock thermalization rather than cascade dissipation of the merger-driven turbulence considered here, with cascade dissipation becoming competitive only beyond ${\sim}0.3$--$0.5\,R_{500c}$ where radiative losses have declined by orders of magnitude. Reconciling this turbulent heating budget with a directly identified shock-dominated heating channel is plausible if AGN-driven turbulence, injected on the smaller bubble/jet-lobe scales discussed above rather than at $r$, itself constitutes a non-negligible share of the total turbulent heating budget in the core; disentangling the relative contributions of shock thermalization and turbulent cascade dissipation, at their respective physically appropriate injection scales, is a natural target for future work. This is also consistent with \citet{shi_etal20}, who demonstrate using Lagrangian tracer particles in non-radiative simulations that turbulence dissipation dominates ICM heating inside $R_{500c}$ following merger events, with the heating rate declining progressively as the merger-driven turbulence dissipates over Gyr timescales.

\begin{figure}
    \begin{center}
    \includegraphics[width=0.45\textwidth]{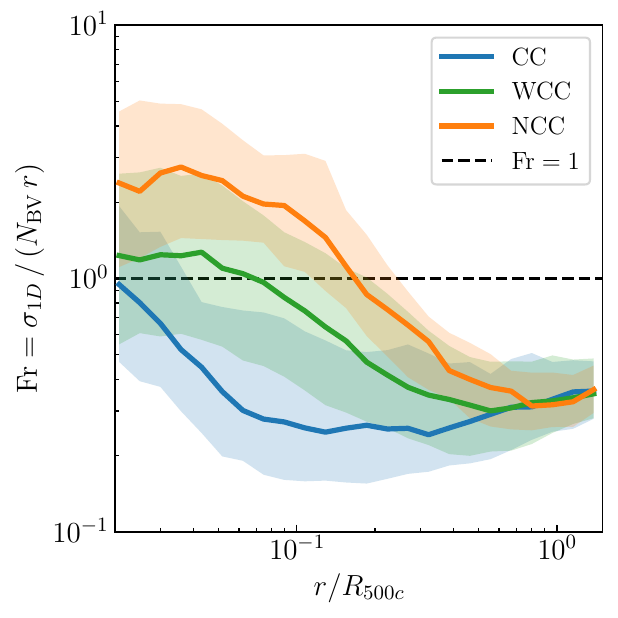}
    \end{center}
    \caption{The Froude number $\mathrm{Fr}$, plotted as a function of radius for different cool-core states (CC: blue; WCC: green; NCC: orange). The dashed line indicates $\mathrm{Fr} = 1$, where turbulence balances buoyancy; buoyancy dominates when $\mathrm{Fr} < 1$, and turbulence dominates when $\mathrm{Fr} > 1$. 
    %{Alt text: Comparison of the Froude number radial profile for CC, WCC, and NCC clusters.}
    }
    \label{fig:tng_froude_cc}
\end{figure}

\subsubsection{ICM Stratification and Buoyancy}

One proposed explanation for low velocity dispersions in cool-core clusters is the suppression of turbulence by buoyancy in the strongly stratified ICM, which can convert turbulent motions into internal gravity waves, thereby reducing the efficiency of isotropic mixing \citep{Shi2018, Zhang2018, Shi_Zhang2019}.

We examine the role of buoyancy versus turbulence in the TNG-Cluster simulations. Figure~\ref{fig:tng_froude_cc} shows the radial profile of the Froude number, defined as
\begin{equation}
\mathrm{Fr} \equiv \frac{\sigma_{\rm 1D}}{rN_{\rm BV}},
\end{equation}
where $N_{\rm BV}$ is the Brunt--V\"{a}is\"{a}l\"{a} frequency (i.e., the inverse of the buoyancy timescale):
\begin{equation}
N_{\rm BV} \equiv \sqrt{\frac{g}{\gamma}\frac{d\ln K}{dr}},
\end{equation}
where $g$ is the gravitational acceleration, $\gamma = 5/3$ is the adiabatic index, and $K$ is the gas entropy. As with the heating-rate estimate above, this definition takes $r$ as the relevant length scale, which is the appropriate choice for large-scale, merger-driven motions but may overestimate the true Froude number for any AGN-driven component of the turbulence injected on smaller, bubble-sized scales; we do not attempt to isolate the two contributions here; and note this as an additional caveat on the quantitative Froude number values, though not on their qualitative radial and cool-core trends. The CC population exhibits $\mathrm{Fr}<1$ throughout most of the central region, reaching values of $\mathrm{Fr}\sim 0.2$--$0.4$ at $r\lesssim 0.1\,R_{500c}$, indicating that gas motions are strongly influenced by buoyancy forces. In this regime, turbulent eddies are unable to efficiently overcome the background entropy stratification, leading to anisotropic motions and the excitation of internal gravity waves. In contrast, NCC clusters maintain $\mathrm{Fr}\gtrsim1$ over a much larger radial range, up to $\sim 0.2\,R_{500c}$, implying that turbulent inertia dominates over buoyancy and allows more isotropic gas motions. The low Froude numbers found in simulated CC clusters demonstrate that buoyancy suppression is already an important component of the gas dynamics in cluster cores. This suggests that any discrepancy in the low-velocity dispersions measured by XRISM/Resolve cannot be attributed solely to stratification. 

\subsubsection{Dependence on Cluster Formation History}\label{sec:formation_history}

\begin{figure}
    \begin{center}
    \includegraphics[width=0.49\textwidth]{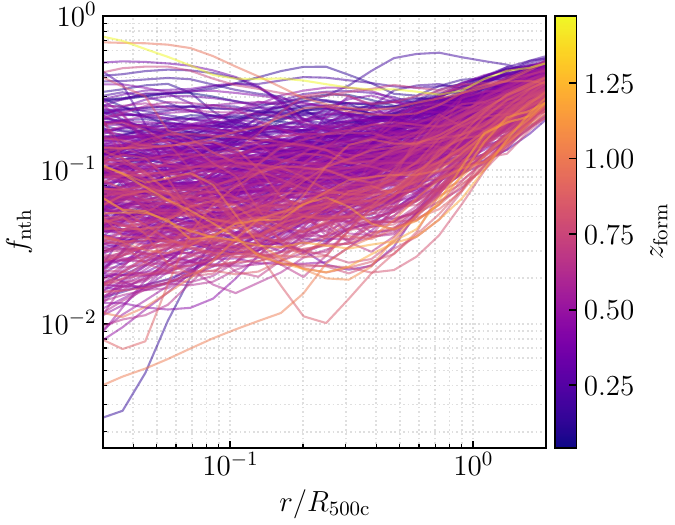}
    \end{center}
    \caption{
     Non-thermal pressure fraction profile for every TNG-Cluster halo, as a function of scaled radius $r/R_{500c}$, color-coded by formation redshift $z_{\rm form}$, defined as the redshift at which the cluster attained half of its present-day mass.
    % {Alt text: Non-thermal pressure fraction profiles for all simulated clusters plotted against radius normalized by R500c. Curves are color-coded by formation redshift.}
     }
    \label{fig:fnt_profile_zform}
\end{figure}

\begin{figure*}
    \begin{center}
    \includegraphics[width=0.99\textwidth]{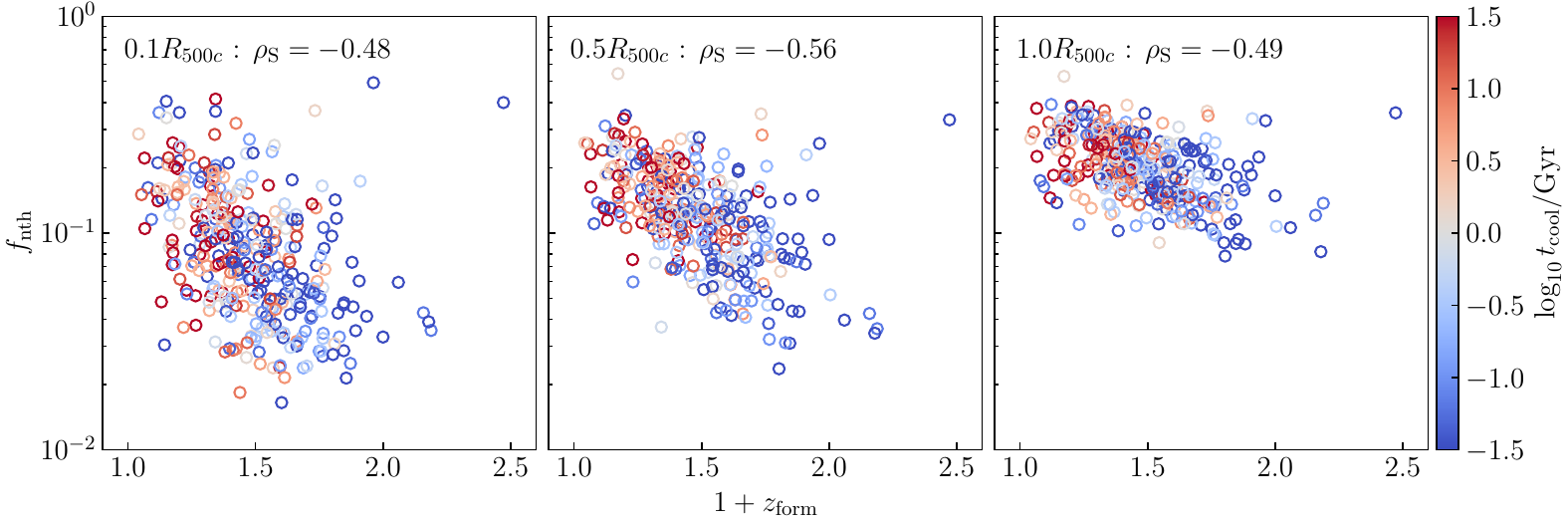}
    \end{center}
    \caption{The non-thermal pressure fraction measured at three different radii: $0.1R_{500c}$ (left panel), $0.5R_{500c}$ (middle panel), and $1.0R_{500c}$ (right panel), plotted as a function of halo formation redshift $1+z_{\rm form}$. The color indicates the central cooling time $t_{\rm cool}$. The plots show that the non-thermal pressure fraction is generally anti-correlated with formation redshift. We also include the value of the Spearman correlation coefficient $\rho_s$ in each panel.
    %{Alt text: Scatter plots of non-thermal pressure fraction versus cluster formation redshift at 0.1, 0.5, and 1.0 R500c. Points are colored by central cooling time.}
    }
    \label{fig:fnt_zform_correlation}
\end{figure*}

Figure~\ref{fig:fnt_profile_zform} shows the non-thermal pressure fraction profiles for individual clusters as a function of formation redshift, $z_{\rm form}$, defined as the redshift at which the cluster assembled half of its present-day mass.

At fixed radius, late-forming clusters (low $z_{\rm form}$) exhibit systematically higher $f_{\rm nth}$ than early-forming systems. This trend persists over the full radial range and reflects the higher level of residual gas motions in dynamically younger clusters. This interpretation is supported by \citet{zhang_etal25}, who show using Lagrangian tracer particles that ICM pair dispersion rates are dramatically elevated during major mergers and decline toward quiescent levels afterward, establishing that a cluster's dynamical history directly governs its present-day ICM transport properties and the amplitude of residual gas motion. Early-forming systems show lower non-thermal pressure support, consistent with more relaxed conditions.

Despite significant scatter among individual halos, the radial trend of increasing $f_{\rm nth}$ is robust. When scaled by $R_{200m}$, the profiles exhibit an approximately self-similar form, particularly at large radii. Deviations are more pronounced in the inner regions, where baryonic processes and dynamical state introduce additional complexity.

Figure~\ref{fig:fnt_zform_correlation} shows more clearly the anti-correlation between $1+z_{\rm form}$ and $f_{\rm nth}$ measured at three different radii: $0.1R_{500c}$, $0.5R_{500c}$, and $1.0R_{500c}$; the Spearman correlation coefficients are $\rho_s = -0.48$, $-0.56$, and $-0.49$, respectively. This anti-correlation is qualitatively consistent with the analytical framework of \citet{shi_etal14, shi_etal15}, which predicts that non-thermal pressure support is sustained by accretion-driven turbulence and should therefore anti-correlate with the formation epoch: clusters that assembled earlier have dissipated their merger-driven non-thermal energy budget over longer timescales. We also color-code the simulation data points by the central cooling time $t_{\rm cool}$. Clusters with shorter $t_{\rm cool}$ generally have higher $z_{\rm form}$ and lower $f_{\rm nth}$. 

While this anti-correlation is moderate rather than strong, it has implications for interpreting the XRISM/Resolve observation of A2029N itself. Deep \textit{Chandra} observations from \citet{watson_etal2026} trace the ICM structures in A2029N and found one of the longest continuous sloshing spirals yet observed due to a subcluster's second core passage roughly $4\,\mathrm{Gyr}$ after the onset of a merger. This indicates that A2029's last major merger occurred well in the past and that the cluster has had a correspondingly long time to relax, consistent with A2029N being a genuinely early-forming system in the sense probed by Figure~\ref{fig:fnt_zform_correlation}, and therefore its low $f_{\rm nth}$ is an expected consequence of its formation history. A direct, quantitative test e.g., reconstructing A2029's assembly history from its merger record and comparing it to the $z_{\rm form}$-$f_{\rm nth}$ relation calibrated here, would be a natural target for future work.

\subsubsection{Non-thermal Pressure Fraction Model Fit}

We model the non-thermal pressure fraction (computed using mass-weighted velocity and temperature profiles) as a function of radius using a phenomenological form that incorporates two physically motivated scales, $R_{500c}$ and $R_{200m}$, in a two-scale exponential model:
\begin{eqnarray}\label{eq:fnt_model}
f_{\rm nth}(r) &=&
f_0 - A_{\rm nth}\left[1-\,\exp\!\left(-\left(\frac{r}{aR_{500c}}\right)^{\alpha}\right)\right] \\ \nonumber
&+&(f_\infty -f_0 + A_{\rm nth})\left[1 - \exp\!\left(-\left(\frac{r}{bR_{200m}}\right)^{\beta}\right)\right],
\end{eqnarray}
where $f_0 \equiv f_{\rm nth}(r \to 0)$ is the central value; $A_{\rm nth}$ characterizes the drop (or rise) in $f_{\rm nth}$ from the cluster center to the edge of the inner cluster region, with a positive (negative) value indicating a drop (rise); $f_\infty$ is the asymptotic value at large radii; $a$ and $b$ are dimensionless scale parameters that represent the size of the inner and outer cluster regions, respectively; and $\alpha$ and $\beta$ control the sharpness of the inner and outer transitions. The first two terms describe the behavior of $f_{\rm nth}$ in the core region of the cluster due to AGN feedback, while the last term describes the behavior of the outskirts due to mergers and accretion. In the last term, we normalize the radius by $R_{200m}$, as the outskirts $f_{\rm nth}$ scales self-similarly to the outer accretion shock radius that is closely tracked by $R_{200m}$ \citep{lau_etal15, aung_etal21}.

Unlike single power-law models \citep{shaw_etal10, vazza_etal18, angelinelli_etal20} or a single exponential model \citep{nelson_etal14b}, this model is constructed to capture the wide range of behavior in both cluster cores and outskirts across different dynamical and cool-core states.

The left panel of Figure~\ref{fig:fnt_profile_cc} shows the best-fit model profiles for the different cool-core bins. The model captures the upturn in $f_{\rm nth}$ in the cores of CC clusters. We also include a comparison to our previous model from \cite{nelson_etal14b}, calibrated against cosmological simulations of galaxy clusters with no radiative cooling and feedback physics, as well as the fit to another non-radiative cosmological simulation from \cite{angelinelli_etal20}. This shows that our new model better recovers the behavior of $f_{\rm nth}$ in cluster cores across different cool-core states. Table~\ref{tab:cc_wcc_ncc_bestfit} lists the corresponding best-fit parameters for the three cool-core bins and for the whole cluster sample, which would be useful for comparing with other simulations or observations of $f_{\rm nth}$ as a function of radius for different cool-core clusters.  

\begin{table}
    \tbl{Best-fit parameters for the median non-thermal pressure fraction profile model (Equation~\ref{eq:fnt_model}) for the CC, WCC, and NCC subsamples, and for all clusters.}{
        \begin{tabular}{lccccccc}
        \hline
        Sample & $f_0$ & $f_{\infty}$ & $A_{\rm nth}$ & $a$ & $b$ & $\alpha$ & $\beta$ \\
        \hline
        CC & $0.103$ & $0.48$ & $0.033$ & $0.041$ & $0.70$ & $3.71$ & $1.89$ \\
        WCC & $0.054$ & $0.49$ & $-0.076$ & $0.37$ & $0.74$ & $0.72$ & $2.20$ \\
        NCC & $0.051$ & $0.53$ & $-0.13$ & $0.29$ & $0.78$ & $0.74$ & $2.39$ \\
        All & $0.067$ & $0.50$ & $-0.017$ & $0.16$ & $0.75$ & $2.51$ & $2.27$ \\
        \hline
        \end{tabular}
    }\label{tab:cc_wcc_ncc_bestfit}
\end{table}

\subsection{Mock Map Analyses}
\label{sec:mock_analyses}

\begin{figure}
    \begin{center}
    \includegraphics[width=0.48\textwidth]{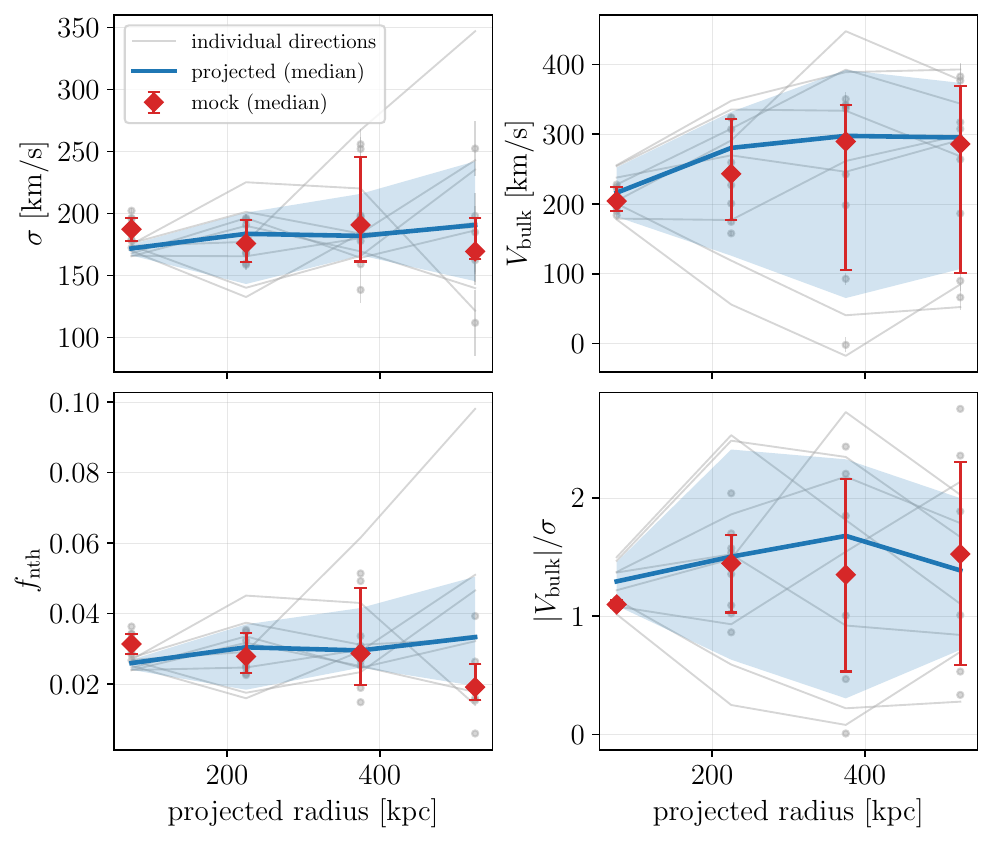}
    \end{center}
    \caption{Profiles of velocity dispersion $\sigma$ (top left), bulk velocity $V_{\rm bulk}$ (top right), the turbulence-only non-thermal pressure fraction $f_{\rm nth}$ (bottom left), and the bulk-to-turbulent velocity ratio $|V_{\rm bulk}|/\sigma$ (bottom right), for the 8 azimuthal directions of a cool-core cluster with HaloID~20, projected along the `x' direction of the simulation. Thin gray lines and points show the individual azimuthal directions, measured directly from the projected maps (lines) and recovered from mock XRISM/Resolve observations (points), respectively; the bold blue line and shaded band show the median and 16th--84th percentile range of the projected values across the 8 directions, and the bold red diamonds with error bars show the same for the mock-recovered values. 
    %{Alt text: Four-panel plot of velocity dispersion, bulk velocity, non-thermal pressure fraction, and the bulk-to-turbulent velocity ratio for eight azimuthal directions in a representative cool-core cluster, each showing individual directions as thin gray traces with a bold median summary overlaid for both the projected and mock-recovered values.}
    }\label{fig:CL20_mock_azu_profiles}
\end{figure}

\begin{figure*}
    \begin{center}
    \includegraphics[width=0.75\textwidth]{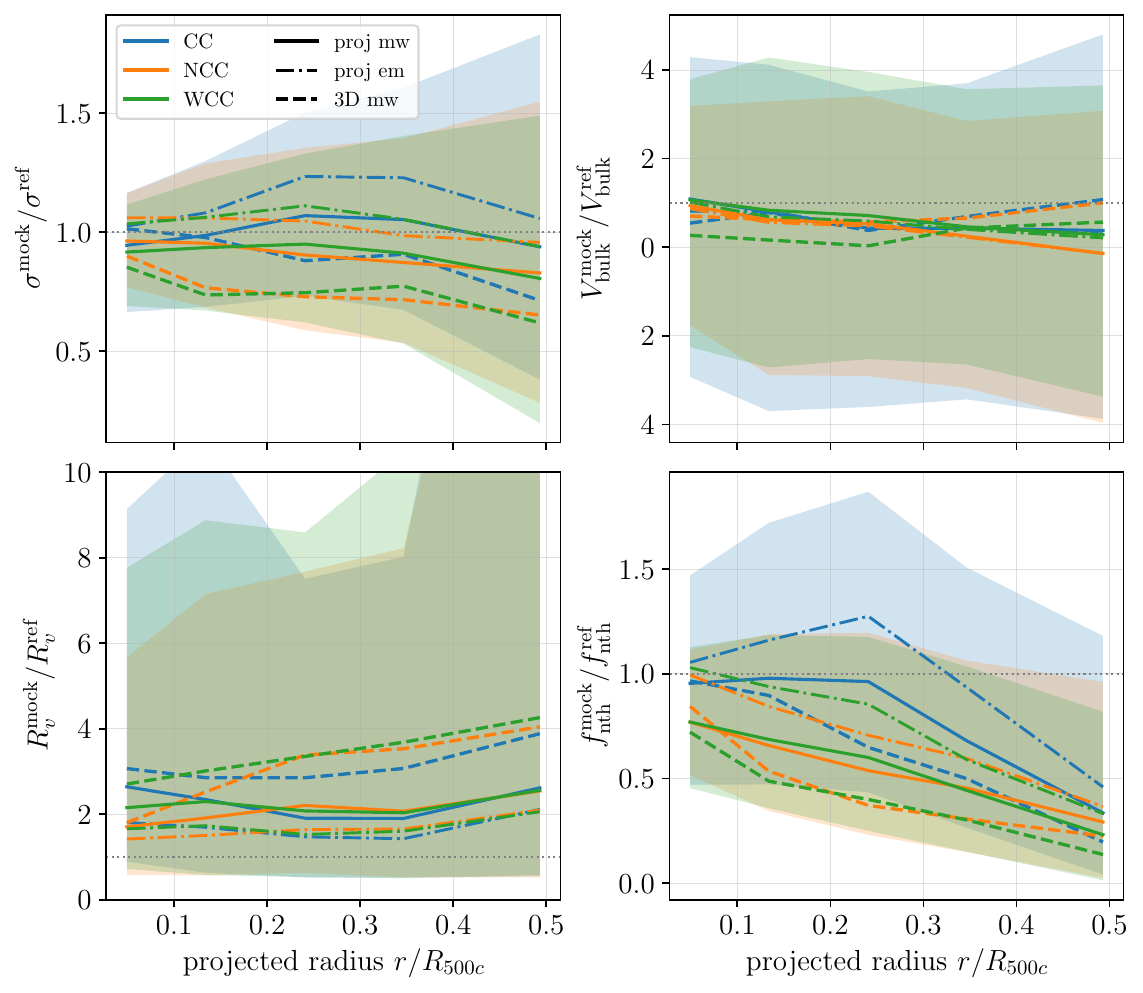}
    \end{center}
    \caption{
    Ratio of the mock-recovered to a reference value, $X^{\rm mock}/X^{\rm ref}$, as a function of projected radius $r/R_{500c}$, for the line-of-sight velocity dispersion $\sigma$ (top left), bulk velocity $V_{\rm bulk}$ (top right), the bulk-to-turbulent velocity ratio $R_v \equiv |V_{\rm bulk}|/\sigma$ (bottom left), and the turbulence-only non-thermal pressure fraction $f_{\rm nth}$ (bottom right), for all forward-modeled mock map pointings, separated by cool-core class (color: CC, blue; WCC, green; NCC, orange). Three reference values are shown for each quantity (line style, black in the legend): the mass-weighted projected value (solid, ``proj mw''), the emission-weighted projected value (dash-dot, ``proj em''), and the intrinsic three-dimensional, spherically-averaged mass-weighted value (dashed, ``3D mw''). Shaded bands show the 16th--84th percentile range of the mock-to-proj-mw ratio; the dotted horizontal line marks unity (perfect agreement).
    %{Alt text: Four-panel plot of the ratio between mock-recovered values and three reference quantities -- mass-weighted projected, emission-weighted projected, and three-dimensional spherically-averaged mass-weighted -- for velocity dispersion, bulk velocity, the bulk-to-turbulent velocity ratio, and the non-thermal pressure fraction, each as a function of projected radius, separated into cool-core, weak cool-core, and non-cool-core classes with shaded percentile bands and a unity reference line.}
    }
    \label{fig:fm_bias_radius}
\end{figure*}

\subsubsection{Mock X-ray values vs Simulation values}
\label{sec:mock_vs_sim}

We use the \texttt{bapec} model within \texttt{XSPEC} to fit each mock XRISM/Resolve spectrum ($E\in[3.0,9.0]$~keV), fitting all parameters except $N_H$, which is fixed to its mock-generation value. The projected bulk velocity is $V_{\rm bulk}=c(z_{\rm fit}-z_{\rm cl})/(1+z_{\rm cl})$, with $z_{\rm cl}=0.0787$. Following the convention in \cite{xrism_a2029_2}, for a single line of sight, the non-thermal pressure fraction from the mock and projected maps is computed as
\begin{equation}
f_{\rm nth} = \frac{\sigma^{2}}{\sigma^{2}+c_{s}^{2}/\gamma}
= \frac{\mathcal{M}^{2}}{\mathcal{M}^{2}+3/\gamma},
\,\, \mathcal{M}\equiv\frac{\sqrt{3}\sigma}{c_{s}},
\label{eq:fnth_los}
\end{equation}
where $\sigma$ is the 1D line-of-sight velocity dispersion, $c_{s}^{2}=\gamma\,k_{B}T/(\mu m_{p})$ is the adiabatic sound speed, and $\gamma=5/3$, $\mu=0.59$. The temperature $T$ is the spectrally-fitted projected temperature. When a coherent line-of-sight bulk motion $V_{\rm bulk}$ is included, the corresponding fraction is
\begin{equation}
f_{\rm nth}^{\rm turb+bulk} = \frac{\mathcal{M}_{v}^{2}}{\mathcal{M}_{v}^{2}+3/\gamma},\, \, \mathcal{M}_{v}\equiv\frac{\sqrt{3\sigma^{2}+V_{\rm Bulk}^{2}}}{c_{s}}.
\label{eq:fnthbulk_los}
\end{equation}

In addition, we also compare the bulk-to-turbulent velocity ratio, defined as
\begin{equation}
R_v \equiv |V_{\rm Bulk}|/\sigma,
\end{equation} 
between mock and simulated values. 

Before comparing the mock maps with A2029, we first check whether it introduces its own bias relative to intrinsic values, using a single representative cluster and the full forward-modeled population. Figure~\ref{fig:CL20_mock_azu_profiles} compares, for HaloID~20 (a cool-core cluster with a sloshing pattern similar to A2029, see Figure~\ref{fig:projected_maps}), the projected and mock-recovered $\sigma$, $V_{\rm bulk}$, $f_{\rm nth}$, and $R_v \equiv |V_{\rm bulk}|/\sigma$ for patches in 8 azimuthal directions as shown in Figure~\ref{fig:xrism_mock_map}. All quantities $\sigma$, $V_{\rm bulk}$, $f_{\rm nth}$, and $R_v$ show little systematic offset in the median values, and the scatter values show good agreement regardless of direction and projected radius. 

Figure~\ref{fig:fm_bias_radius} extends this to the full forward-modeled population, pooled and split by cool-core class. For each pointing we compare the mock fit to the mass-weighted and emission-weighted projected values in the same $3'\times3'$ box. The same comparison is applied to $V_{\rm bulk}$, $R_v$, and $f_{\rm nth}$.
 
The mock-to-projected (mass-weighted) ratio for $\sigma$ and $f_{\rm nth}$ is close to unity for WCC/NCC at all radii, but the CC clusters show a modest excess peaking at $\sim1.25$ near $r/R_{500c}\sim0.2$--$0.4$, with $16$--$84$th percentile scatter widening from $0.8$--$1.6$ to $0.5$--$2.0$ over the same range. $V_{\rm bulk}$ and $R_v$ show much larger scatter for all classes, reflecting the instability of a ratio that crosses zero rather than a comparable fit bias. The emission-weighted projected comparison follows the same pattern but more strongly: the CC peak is comparable or slightly higher, while WCC/NCC decline further outward ($\sim0.6$--$0.7$ by $r/R_{500c}\sim0.5$). For $R_v$, the mock values recover noticeably better against the emission-weighted reference than the mass-weighted one, consistent with the mock fit itself being an emission-weighted measurement.

We also compare the mock values against the intrinsic 3D spherically-averaged values. To be consistent with the mock and projected map measurements and observations, we use the  $f_{\rm nth}$ definition in Eq.~(\ref{eq:fnth_los}), with 
\begin{equation}
f_{\rm nth} = \frac{\sigma_{\rm 1D}^{2}}{\sigma_{\rm 1D}^{2}+c_{s}^{2}/\gamma},
\end{equation}
where $\sigma_{\rm 1D}$ is the same 1D velocity dispersion in Eq.(\ref{eq:sigma_1d}). 
The mock values generally underestimate the ``true'' 3D values. The mock-to-3D ratio for $\sigma$ is $\sim0.9$ in CC clusters and $\sim0.75$ in WCC/NCC at $r\simeq0.2\,R_{500c}$ ($\sim10\%$ and $\sim25\%$ underestimates). This is counter-intuitive: a sightline at projected radius $r_{\rm proj}$ samples only $r_{\rm 3D}\geq r_{\rm proj}$, and $\sigma_{\rm 1D}(r)$ increases with $r$ here, so projection should raise, not lower, the projected value.

The most likely explanation is that emission weighting restricts the effective sampled path length along a sightline below the turbulence outer scale $r$. Following \citet{zhuravleva_etal12}, the effective length $l_{\rm eff}$ at projected radius $R$ is defined as the region size, centered on the point of closest approach along the line of sight, that contributes half of the total flux along the full sightline:
\begin{equation}
\int_{0}^{l_{\rm eff}} w\!\left(\sqrt{R^{2}+z^{2}}\right) dz
= \frac{1}{2}\int_{0}^{\infty} w\!\left(\sqrt{R^{2}+z^{2}}\right) dz,
\label{eq:leff}
\end{equation}
where $w$ is the weight along the line of sight (the emission weight of Equation~\ref{eq:em_weight}) and $z$ is the line-of-sight coordinate. We replace $\infty$ in the upper limit on the RHS with the projection box length = 3072~kpc. $l_{\rm eff}$ shrinks as the weight becomes more centrally concentrated toward the point of closest approach, and approaches the full sightline length only when $w$ is spatially uniform. For a Kolmogorov-like turbulent cascade, sampling only $l_{\rm eff}<r$ predicts $\sigma^{\rm mock}/\sigma^{\rm 3D}\sim(l_{\rm eff}/r)^{1/3}$. Using our own measured $l_{\rm eff}(r)/r$ profiles, this gives $\sim0.77$ for CC at $r\simeq0.2\,R_{500c}$ ($l_{\rm eff}/r\simeq0.45$), in reasonable agreement with the observed $\sim0.9$. The same scaling under-predicts the WCC/NCC deficit ($l_{\rm eff}/r\simeq0.65$, $0.80$ implies $\sim0.87$, $0.93$ versus the observed $\sim0.75$), suggesting $r$ is not a uniformly appropriate turbulence outer scale across cool-core classes (Section~\ref{sec:heat_cool}). 
%A full identification of the mechanism is left to an upcoming work on ICM velocity power spectra. 

Consequently the mock-to-3D ratio for $f_{\rm nth}$ is $\sim0.8$ in CC and $\sim0.5$ in WCC/NCC ($\sim20\%$ and $\sim50\%$ underestimates). This deficit grows with projected radius (Figure~\ref{fig:fm_bias_radius}, bottom right), driving the declining mock-to-reference trend there: because the true 3D $f_{\rm nth}$ also increases with radius (Section~\ref{sec:cc_status}), the two trends partially cancel, so a flat or declining mock/projected $f_{\rm nth}$ does not imply the true 3D value is flat or declining. The mock-to-3D ratio for $V_{\rm bulk}$ spans $\sim0.1$--$1.0$ with no clear cool-core dependence (an underestimate of a few percent to nearly $90\%$), so $R_v$ is correspondingly overestimated. Dividing the mock-to-3D by the mock-to-projected ratio gives an implied projected-to-3D ratio of $\sim0.72$--$0.75$ for $\sigma$, essentially the same across classes.
 
We attribute the CC-specific $\sigma$ (and $f_{\rm nth}$) excess to the recovery of a single emission-weighted dispersion from a sightline spanning a wider range of temperatures and velocities in cool cores, compounded by declining photon statistics at larger radii. This mechanism is directly supported by \citet{Truong2024}, who used mock XRISM/Resolve spectra of Perseus-like cool-core clusters drawn from the same TNG-Cluster simulation we use here, and showed that hot gas along the line of sight can bias the single-component velocity dispersion inferred from high-energy X-ray lines to values in excess of the true value. This bias is attributed to a combination of projection and multi-temperature, multi-velocity structure along the sightline. The bulk velocity $V_{\rm bulk}$, recovered from the line centroid rather than its width, is less sensitive to this effect.
 
\subsubsection{Azimuthal Variations and Comparison to A2029}

\begin{figure*}
    \begin{center}
    \includegraphics[width=0.99\textwidth]{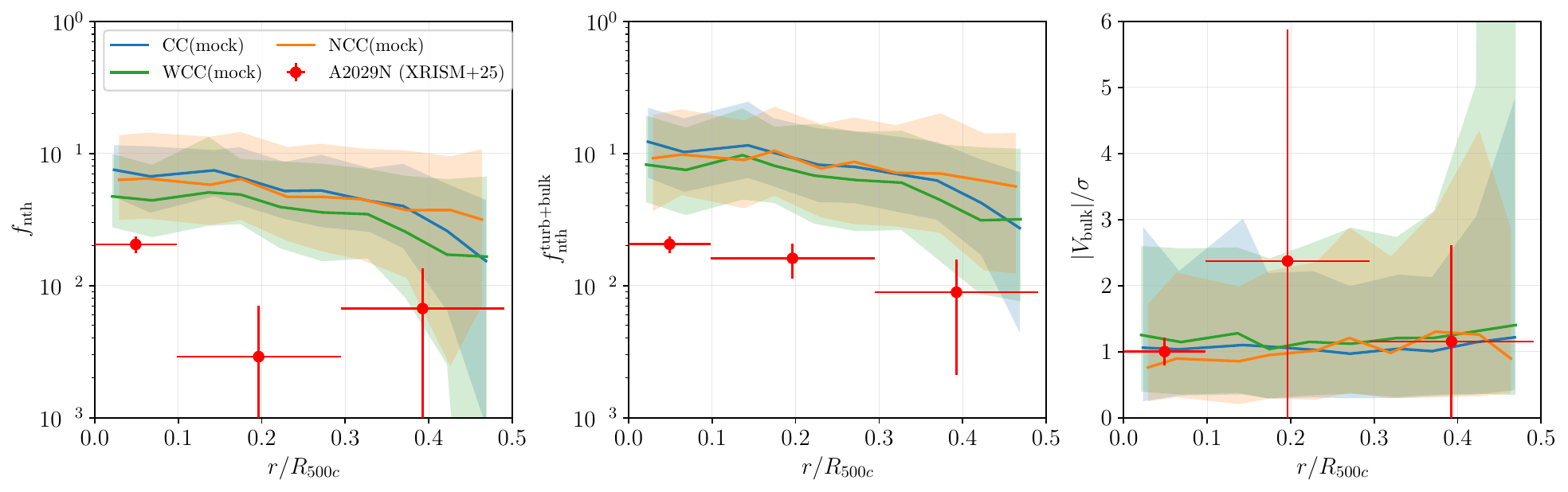}
    \end{center}
    \caption{{\em Left} panel: Non-thermal pressure fraction profiles obtained from analyzing mock XRISM maps for the CC (blue), WCC (green), and NCC (orange) clusters. The solid lines are the median over all three orthogonal projections of all clusters within the bin, and the shaded region covers the 16th--84th percentiles. The red data points are the XRISM/Resolve measurements of A2029N from \cite{xrism_a2029_2}.
    {\em Middle} panel: same as the left panel, but with the non-thermal pressure fraction computed using both LOS velocity dispersion and LOS bulk velocity.
    {\em Right} panel: Ratio of LOS bulk velocity to LOS velocity dispersion.
    {Alt text: Comparison between simulated XRISM-like measurements and XRISM observations of Abell 2029. Left: projected non-thermal pressure fraction computed from turbulent velocity dispersion only. Middle: projected non-thermal pressure fraction including both turbulent and bulk motions. Right: ratio of line-of-sight bulk velocity to velocity dispersion.}}
    \label{fig:azu_fnt_profiles}
\end{figure*}

To place these azimuthal variations in the context of XRISM/Resolve observations, we compare the range of inferred non-thermal pressure fractions in simulated clusters with the measurements of A2029. As shown in Figure~\ref{fig:azu_fnt_profiles}, the azimuthally sampled mock XRISM profiles for CC, WCC, and NCC systems span a broad range at fixed radius, with variations reaching factors of a few in $f_{\mathrm{nth}}$, particularly beyond $r \gtrsim 0.2$--$0.3\,R_{500c}$. For each cool-core state bin, we take the median and 16th--84th percentiles over all 8 azimuthal arms for all three ($x,y,z$) projections.

In the turbulence-only case (left panel of Figure~\ref{fig:azu_fnt_profiles}), the lower envelope of the simulated distribution extends downward toward the XRISM measurements of A2029. Specifically, A2029's innermost point ($r\simeq0.05\,R_{500c}$, $f_{\rm nth}=0.021$) falls at the 1.9th percentile of the simulated CC distribution at that radius; the outermost point ($r\simeq0.4\,R_{500c}$, $f_{\rm nth}=0.0067$) falls at the 6.1st percentile; but the intermediate point ($r\simeq0.2\,R_{500c}$, $f_{\rm nth}=0.003$) falls at the 0.0th percentile, below every simulated CC pointing found at that radius, not merely in the low tail. The result indicates that similarly low inferred turbulent pressure fractions can arise from projection and azimuthal sampling effects even in otherwise typical simulated clusters, though the measurement at the intermediate radius remains an outlier even relative to this scatter.
 
When the mock $f_{\rm nth}$ is instead computed including both the line-of-sight (LOS) velocity dispersion and the bulk velocity (middle panel), the simulated profiles rise substantially relative to the turbulence-only case, though the median mock $f_{\rm nth}^{\rm turb+bulk}$ itself declines with radius for the CC class, from $0.106$ at $r\simeq0.05\,R_{500c}$, to $0.094$ at $r\simeq0.2\,R_{500c}$, to $0.058$ at $r\simeq0.4\,R_{500c}$. Relative to this distribution, A2029's three points fall at the 0.9th, 0.0th, and 2.8th percentiles, respectively, lower than their turbulence-only counterparts. This increase relative to the turbulence-only case reflects the significant coherent bulk motion carried by the simulated clusters, with $R_v \gtrsim 1$ over much of the radial range (right panel). As a consequence, A2029N now lies systematically and substantially below the simulated distribution at all three radii, by factors of a few relative to the lower envelope and by up to an order of magnitude relative to the median, so that including bulk motion widens the discrepancy rather than reducing it. 
 
We note that the bulk-to-turbulent ratio measured for A2029N itself ($R_v\sim1$--$2$) is fully consistent with the simulated range; the disagreement therefore arises from the lower absolute amplitude of both the turbulent and bulk gas motions in A2029, rather than from how its non-thermal pressure budget is partitioned between the two.
 
In both definitions, A2029N lies at or below the lowest values recovered from the mock observations across multiple radii, and most markedly so once bulk motion is included, especially in the outskirts. This suggests that while azimuthal variation and line-of-sight projection can significantly reduce the inferred level of gas motion, they are unlikely to fully account for the most extreme cases. This conclusion, which follows directly from our forward-modeled XRISM comparison, is qualitatively consistent with intrinsic velocity-decomposition studies of TNG-Cluster that also find turbulent motions to be a subdominant component of the central velocity field \citep[e.g.][]{saha_etal26}. A2029N may therefore represent a rare dynamical state with intrinsically suppressed gas motions, or point to missing or inaccurately modeled physical processes in current simulations.
 
\citet{xrism_comparison} reported larger discrepancies between A2029N and the TNG-Cluster simulations. A key methodological difference lies in the sampling strategy. In our analysis, we systematically extracted spectra from all eight azimuthal directions for each $x$, $y$, and $z$ projection of each halo. In contrast, \citet{xrism_comparison} selected only six random locations along two orthogonal directions. Consequently, our line-of-sight sampling is more extensive, reducing statistical variance and averaging over azimuthal fluctuations, which leads to smaller apparent differences between simulations and observations.
 
\section{Discussions}
\label{sec:discussions}

\subsection{Comparison with related TNG-Cluster analyses}
\label{sec:saha_comparison}

A complementary analysis of gas motions in TNG-Cluster was recently presented by \citet{saha_etal26}, who applied a multi-scale filtering Reynolds decomposition to separate the intrinsic velocity field into coherent bulk and small-scale turbulent components. Their analysis and ours are complementary in scope: \citet{saha_etal26} characterize the intrinsic, multi-scale-decomposed three-dimensional velocity field, whereas we forward-model the XRISM/Resolve observable and compare directly with the measurements of Abell~2029. The two approaches are consistent where they overlap. The central upturn in $f_{\rm nth}$ that our two-scale model captures through the $A_{\rm nth}$ term (Section~\ref{sec:cc_status}) corresponds to the central peak in their turbulent velocity dispersion, and our turbulence-only mock profiles reach the similarly low core values they report, with the larger part of the inferred non-thermal motion attributable to coherent bulk flows. The added value of our analysis is to explicitly show how these low intrinsic turbulent fractions, once projected and azimuthally sampled, map onto the observed A2029N measurements. The mock-to-3D comparison of Section~\ref{sec:mock_analyses} quantifies this mapping directly: the forward-modeled $\sigma$ and $f_{\rm nth}$ underestimate their intrinsic 3D, spherically-averaged counterparts by a cool-core-dependent factor ($\sim10\%$ and $\sim20\%$ for CC, $\sim25\%$ and $\sim50\%$ for WCC/NCC, respectively), while $V_{\rm bulk}$ is underestimated by a comparably large but highly variable amount, and the bulk-to-turbulent ratio $R_v$ is correspondingly overestimated, which is a forward-modeling-specific systematic that is not present in, and cannot be recovered from, an intrinsic 3D decomposition such as \citet{saha_etal26}.

\subsection{Limitations of the Current Study}

While our analysis provides a physically motivated interpretation of XRISM measurements within the TNG-Cluster simulations, several limitations should be kept in mind. First, the IllustrisTNG model represents only one realization of subgrid physics for galaxy formation and feedback. In particular, the treatment of AGN feedback, including the balance between kinetic and thermal modes, plays a central role in regulating gas motions in cluster cores. Recent high-resolution simulations tailored to XRISM observations indicate that neither AGN feedback nor sloshing motions alone can reproduce the observed velocity structure of Perseus, suggesting that the coupling between feedback, stratification, and large-scale cluster dynamics remains incompletely understood \citep{Bellomi2025}.

Different implementations of AGN feedback in other simulation suites, as shown by \citet{xrism_comparison} and FLAMINGO \citep{flamingo}, can produce systematically different levels of turbulence and bulk flows in the ICM. As such, the level of non-thermal pressure support predicted here is not unique, and part of the tension with XRISM observations may reflect model-dependent uncertainties rather than a fundamental discrepancy. In this context, we emphasize that both the $f_{\rm nth}$ profiles and the best-fit parameters of our two-scale model (Table~\ref{tab:cc_wcc_ncc_bestfit}) are specific to the IllustrisTNG feedback prescription. Their normalization and shape, particularly in the core where AGN feedback dominates, may differ in simulations adopting alternative subgrid models, such as FLAMINGO or Magneticum. A direct comparison of the fitting function across multiple simulation suites would test the robustness of these parameters and establish which features are generic predictions of hierarchical structure formation rather than consequences of a specific feedback implementation.

Second, the resolution of TNG-Cluster, while sufficient for capturing large-scale gas dynamics, does not fully resolve the turbulent cascade in the ICM. Small-scale turbulence and dissipation processes are modeled only implicitly, which may introduce biases in the inferred velocity dispersion and the non-thermal pressure fraction. In particular, unresolved mixing and viscosity could alter both the amplitude and radial structure of gas motions, especially in cluster cores where XRISM measurements are most constraining.

Third, our forward-modeling approach with XRISM/Resolve mock maps, while incorporating instrumental responses and projection effects, does not fully capture the complexity of observational analysis. For example, assumptions about spectral modeling, background treatment, and spatial extraction regions can introduce additional systematic uncertainties that are not explicitly included in our mock analysis. These effects may further broaden the range of inferred gas motions in real observations. In particular, our current mock pipeline does not model the instrumental point spread function (PSF; Section~\ref{sec:mock_data}), which is required to quantify how gas motions in the unobserved regions surrounding a single central pointing leak into that pointing's inferred velocity measurement. Incorporating the PSF is a natural, and currently missing, extension needed to directly quantify this systematic for the many XRISM-observed clusters with only a single central pointing.

\subsection{Future Directions and Observational Prospects with XRISM}

On the theoretical side, a key next step is to extend this analysis to a broader range of simulation models with varying subgrid physics, particularly alternative prescriptions for AGN feedback and turbulence generation. Comparisons across multiple simulation suites, such as FLAMINGO and Magneticum, will be essential for disentangling robust physical predictions from model-dependent effects. Higher-resolution simulations and targeted zoom-in studies will also be critical for resolving the small-scale structure of turbulence and for improving predictions of ICM velocity fields.

On the observational side, expanding the sample of clusters with high-quality XRISM measurements will be crucial. Current constraints are based on a limited number of systems, and the apparent tension with simulations may be influenced by small-number statistics and selection effects. A larger, systematically selected sample spanning a range of masses, dynamical states, and redshifts will enable a more robust characterization of the distribution of ICM gas motions and non-thermal pressure support. In this regard, A2029N may not be unique: recent {XRISM}/Resolve observations of the relaxed cool-core cluster Abell~1795 likewise report low velocity dispersions and a non-thermal pressure fraction falling below $1\%$ at large radius \citep{xrism_a1795}, suggesting that such quiescent ICM conditions may occur in more than one system. In addition, current {XRISM} measurements are limited by the small number of pointings per system, making it difficult to disentangle cluster-to-cluster variations from spatial variations within a single cluster.
Wider spatial coverage with multiple pointings per cluster will be important for robustly characterizing the radial and azimuthal dependence of ICM gas motions; further XRISM/Resolve observations of A2029 with additional pointing arms are anticipated and will directly test whether the extreme percentile ranking reported here persists with more complete azimuthal coverage. Additional observations of cluster outskirts and dynamically relaxed systems will help determine whether extreme low-velocity cases such as Abell~2029 are rare outliers or indicative of missing physics.

Ultimately, progress will require a joint effort that combines improved simulations, detailed forward modeling, and expanded observational datasets. Such an approach will enable the use of ICM gas motions as a precision probe of structure formation and feedback processes in galaxy clusters.

\section{Conclusions}
\label{sec:summary}

We investigated the origin and observational inference of non-thermal pressure support in the ICM using the TNG-Cluster simulation to interpret recent XRISM/Resolve measurements of gas motions. Our main conclusions are:

\begin{itemize}

\item The non-thermal pressure fraction $f_{\mathrm{nth}}$ increases robustly with radius. Cool-core systems exhibit reduced central gas motions compared with non-cool-core systems, and late-forming clusters retain enhanced support across all radii. We provide a two-scale fitting function that captures both the inner suppression and the outer rise of $f_{\mathrm{nth}}$ across cool-core states.

\item Turbulent heating in CC-cluster cores is of the same order as radiative cooling ($Q_{\rm heat}/Q_{\rm cool}\sim1$--$2$ near $r\simeq0.05\,R_{500c}$, with scatter spanning both sides of unity); given the factor-of-a-few uncertainty in the heating-rate prefactor $\eta$ (Section~\ref{sec:heat_cool}), the true ratio could plausibly fall below unity. WCC/NCC clusters, by contrast, show heating well in excess of cooling ($\sim5$--$50$) at all radii probed, a conclusion robust to this same uncertainty. This does not identify turbulent cascade dissipation as the dominant CC heating channel, and given the uncertainty in $\eta$, it does not rule out an energetic shortfall there either; for WCC/NCC, however, $Q_{\rm heat}/Q_{\rm cool}$ remains well above unity even under this same uncertainty, ruling out an energetic shortfall there. The systematically low Froude numbers ($\mathrm{Fr}\lesssim0.5$) in cool cores further suggest the XRISM/Resolve tension is unlikely to be resolved by stratification alone, though this estimate shares the same large-injection-scale assumption as the heating-rate calculation above. 

\item The mock-recovered velocity dispersion $\sigma$ underestimates the true intrinsic 3D spherically-averaged values by $\sim10\%$ for CC and $\sim25\%$ for WCC/NCC systems, and for the non-thermal pressure fraction $f_{\rm nth}$ by $\sim20\%$ and $\sim50\%$ respectively, at $\sim 0.2 R_{500c}$. The underestimation increases with radius, thereby partially canceling the radially increasing trend in intrinsic 3D $f_{\rm nth}$. Thus, a flat or declining XRISM-observed 
$\sigma$ or $f_{\rm nth}$ profile, as seen in A2029, does not necessarily imply a flat or declining intrinsic 3D profile.
 
\item In the turbulence-only definition of $f_{\rm nth}$, the lower envelope of simulated cool-core clusters can approach, but not fully reach, the A2029N measurements. Including coherent bulk motions widens the discrepancy instead. Even with the full range of azimuthal variation and projection scatter, the observed level of $f_{\rm nth}$ cannot be accounted for with bulk motions included for A2029N. Quantitatively, A2029N's three measured points fall at the <6th percentile of the simulated CC distribution in the turbulence-only definition, and at the <3rd percentile once bulk motion is included, consistently in or below the extreme low tail of the simulated population, not merely below its median. As the observed bulk-to-turbulent ratio of A2029N is consistent with the simulations, the tension reflects the unusually low absolute amplitude of its gas motions rather than a different partition between bulk and turbulent components.

\end{itemize}

The remaining tension points to either rare dynamical states with intrinsically suppressed gas motions or to missing or inaccurately modeled physics, particularly in AGN feedback and core gas dynamics. Our work complements analyses of the intrinsic velocity field in TNG-Cluster \citep{Ayromlou2024, saha_etal26} by linking that structure to the observational biases governing its inference. Future high-throughput X-ray microcalorimeter missions, such as {NewAthena}, and the Line Emission Mapper or its re-incarnation, with wider spatial coverage, will be essential to separate bulk flows, turbulent broadening, and projection effects, and thereby to establish whether systems like A2029N are rare outliers or indicate a broader shortcoming of current ICM models.

\begin{ack}
The authors thank the referee for helpful feedback and suggestions, and Nhut Truong, Congyao Zhang, and John ZuHone for useful comments. This work was supported in part by the Fund for the Promotion of Joint International Research, JSPS KAKENHI Grant Number 25K01026 (EL, NO). NO acknowledges partial support from the Organization for the Promotion of Gender Equality at Nara Women's University.

This work uses publicly available data from the TNG-Cluster simulation. The TNG-Cluster simulation has been run on several computer clusters: as part of the TNG-Cluster project on the HoreKa supercomputer, funded by the Ministry of Science, Research and the Arts Baden-W\"{u}rttemberg and by the Federal Ministry of Education and Research; the bwForCluster Helix supercomputer, supported by the state of Baden-W\"{u}rttemberg through bwHPC and the German Research Foundation (DFG) through grant INST 35/1597-1~FUGG; the Vera, Cobra, and Raven clusters of the Max Planck Computational Data Facility (MPCDF); and the BinAC cluster, supported by the High Performance and Cloud Computing Group at the Zentrum f\"{u}r Datenverarbeitung of the University of T\"{u}bingen, the state of Baden-W\"{u}rttemberg through bwHPC and the German Research Foundation (DFG) through grant no.\ INST 37/935-1~FUGG. Analysis has been carried out on the Vera supercomputer of the Max Planck Institute for Astronomy (MPIA).
\end{ack}

% \section*{Funding}
%  This research was supported by ...

% \section*{Data availability}
%  The data underlying this article are available ...

\bibliographystyle{pasj}
\bibliography{references}

\appendix

\section*{Comparison between mass-weighted and emission-weighted profiles}\label{sec:mw_vs_sp}

\begin{figure*}
    \begin{center}
    \includegraphics[width=0.99\textwidth]{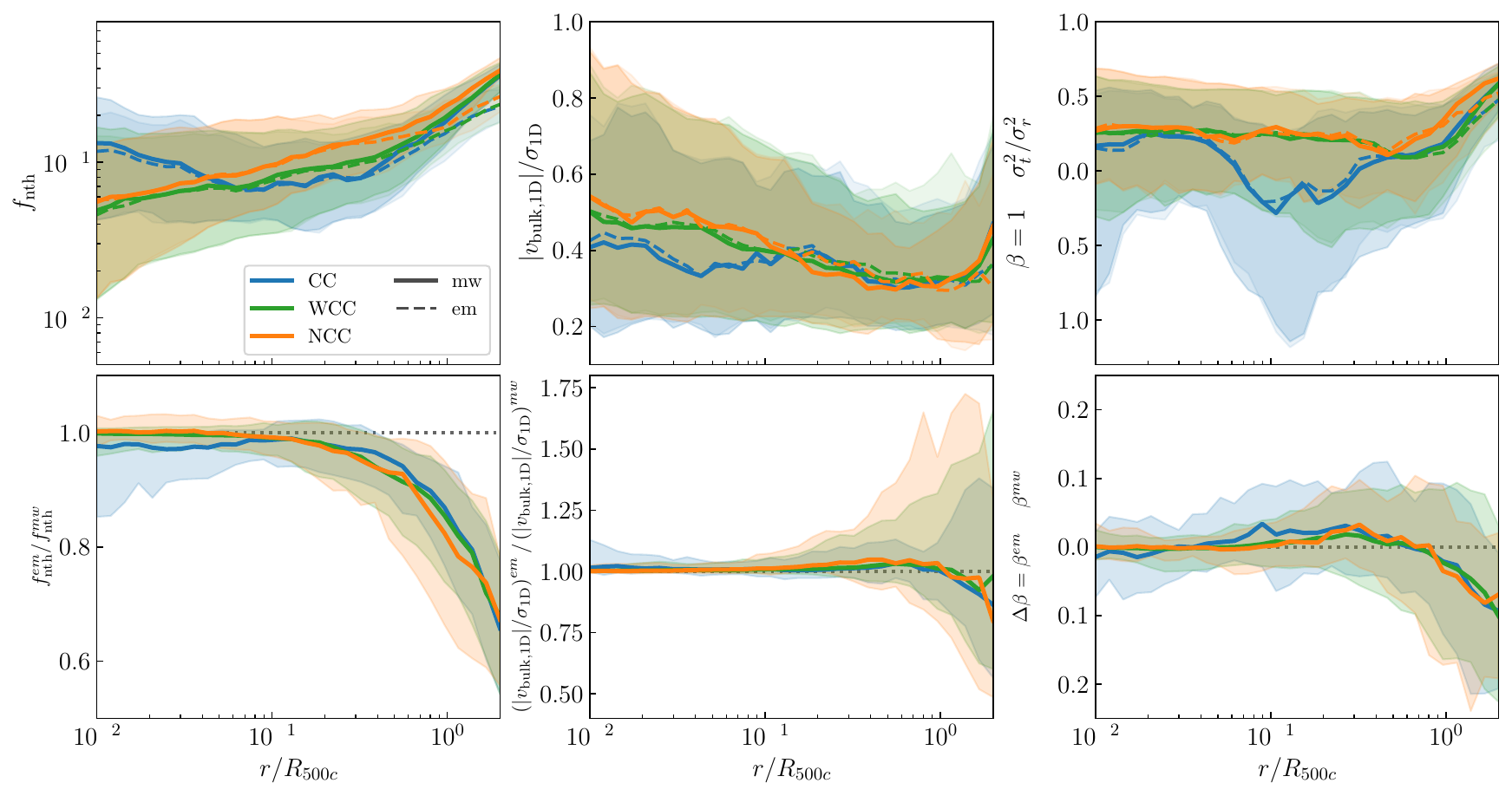}
    \end{center}
    \caption{Comparison of mass-weighted (solid lines) and emission-weighted (dashed lines) kinematic profiles for the 352 TNG-Cluster halos at $z=0$, stratified by cool-core class: cool-core (CC, blue), weak cool-core (WCC, green), and non-cool-core (NCC, orange). Shaded bands enclose the 16th--84th percentile scatter across halos in each class.
    \textit{Top row:} Profiles of the non-thermal pressure fraction $f_\mathrm{nth}$ (\textit{left}), the radial bulk-streaming ratio $|v_r|/\sigma_{1\mathrm{D}}$ (\textit{center}), and the velocity anisotropy parameter $\beta = 1 - \sigma_t^2/\sigma_r^2$ (\textit{right}), as a function of scaled radius $r/R_{500c}$. \textit{Bottom row:} The ratio of emission-to-mass-weighted $f_\mathrm{nth}$ (\textit{left}) and $|v_r|/\sigma_{1\mathrm{D}}$ (\textit{center}), and the difference in anisotropy $\Delta\beta = \beta^\mathrm{em} - \beta^\mathrm{mw}$ (\textit{right}). Dotted horizontal lines mark unity (ratio panels) and zero ($\Delta\beta$ panel). 
    %{Alt text: Comparison of mass-weighted and emission-weighted kinematic profiles for cool-core classes. Top row shows non-thermal pressure fraction, bulk-streaming ratio, and velocity anisotropy as functions of radius. Bottom row shows ratios or differences between the two weighting schemes.}
    }\label{fig:mw_vs_spec}
\end{figure*}

Figure~\ref{fig:mw_vs_spec} compares the mass-weighted (solid lines) and emission-weighted (dashed lines) profiles of $f_\mathrm{nth}$, $|v_{\rm bulk}|/\sigma_{1\mathrm{D}}$, and $\beta$ as a function of $r/R_{500c}$ for the three cool-core classes, together with the corresponding ratio and difference panels (bottom row).

For the non-thermal pressure fraction $f_\mathrm{nth}$ (left column), the two weightings agree to within a few per cent at all radii beyond $r \gtrsim 0.1\,R_{500c}$, with the emission-to-mass-weighted ratio remaining close to unity across all cool-core classes (bottom-left panel). At smaller radii the ratio falls slightly below unity, indicating that the X-ray emissivity weighting down-weights the cool, low-emissivity gas that contributes disproportionately to the mass-weighted turbulent pressure in cluster cores.

The radial streaming ratio $|v_{\rm bulk}|/\sigma_{1\mathrm{D}}$ shows a more pronounced discrepancy: the emission-weighted values are systematically suppressed relative to the mass-weighted values at $r \lesssim 0.3\,R_{500c}$, reflecting the fact that cool, infalling gas in the core, which carries substantial radial bulk momentum relative to the halo center, contributes little X-ray emission and is therefore de-weighted in the emission weighting scheme. The effect is strongest for NCC clusters, which at intermediate radii $0.1 \lesssim r/R_{500c} \lesssim 0.5$ show a ratio $(|v_{\rm bulk}|/\sigma)^{\rm em}/(|v_{\rm bulk}|/\sigma)^{\rm mw}$ exceeding unity by up to $\sim$30--50\,\%, indicating that X-ray-bright gas in this regime moves preferentially along the radial direction compared to the full particle population.

These weighting-dependent offsets can be compared with recent MHD simulation results. E.g. \citet{vazza_brunetti26} reported that the emission-weighted velocity dispersion on average is $\sim 30\%$ smaller than the volume-weighted turbulent velocities; we find a much smaller difference $\lesssim 5\%$ between mass-weighted and emission-weighted dispersions in cluster cores.  We note that the two comparisons are not identical: \citet{vazza_brunetti26} contrast volume- and emission-weighted quantities, whereas our comparison is between mass- and emission-weighted quantities. Since volume weighting gives a greater relative weight to low-density gas than mass weighting does, the volume-to-emission contrast is expected to be the larger of the two, and the two results should be interpreted with this difference in mind.

For the velocity anisotropy profile, the difference between the mass-weighted and emission-weighted profiles is $\Delta\beta = \beta^\mathrm{em} - \beta^\mathrm{mw} \approx +0.1$ -- $0.2$ for the CC clusters at $r < 0.1\,R_{500c}$, while the WCC and NCC clusters display $\Delta\beta \approx 0$ at all radii.

Taken together, these results demonstrate that $f_\mathrm{nth}$ is largely robust to the choice of kinematic weighting, while the inferred bulk streaming fraction and velocity anisotropy can differ substantially between mass-weighted and emission-weighted estimates in cluster cores, with the sign and magnitude of the bias strongly correlated with cool-core state. These systematic offsets must be accounted for when interpreting XRISM velocity measurements of relaxed, cool-core clusters.

\end{document}